\documentclass[useAMS,usenatbib]{mn2e}

\usepackage{graphicx}
\usepackage{times}
\usepackage{natbib}
\usepackage{amssymb,amsfonts,txfonts}
\usepackage[latin1]{inputenc}

\newif\ifAMStwofonts
\AMStwofontstrue

\renewcommand{\vec}[1]{\bmath{#1}}
\newcommand{\be}{\begin{equation}}
\newcommand{\ee}{\end{equation}}
\newcommand{\ba}{\begin{eqnarray}}
\newcommand{\ea}{\end{eqnarray}}
\newcommand{\brr}{\begin{array}}
\newcommand{\err}{\end{array}}
\newcommand{\bc}{\begin{center}}
\newcommand{\ec}{\end{center}}

\newcommand{\msun}{\,h^{-1}M_\odot}

\newcommand{\MpcI}{\,\rm{Mpc}^{-1}}
\newcommand{\fom}{{\rm FoM_{SFT}}}
\newcommand{\fomdetf}{{\rm FoM_{DETF}}}

\newcommand{\fnl}{f_\mathrm{NL}}

\newcommand{\mincir}{\raise
  -2.truept\hbox{\rlap{\hbox{$\sim$}}\raise5.truept \hbox{$<$}\ }}
\newcommand{\magcir}{\raise
  -2.truept\hbox{\rlap{\hbox{$\sim$}}\raise5.truept \hbox{$>$}\ }}
\newcommand{\siml}{\raise
  -2.truept\hbox{\rlap{\hbox{$\sim$}}\raise5.truept \hbox{$<$}\ }}
\newcommand{\simg}{\raise
  -2.truept\hbox{\rlap{\hbox{$\sim$}}\raise5.truept \hbox{$>$}\ }}

\begin{document}
\title[Cluster constraints on non-Gaussianity]
  {The potential of X-ray cluster surveys to constrain primordial
    non-Gaussianity} 
\author[B.\,Sartoris et al.]{B.\,Sartoris$^{1,2,3}$, S.\,Borgani$^{1,2,3}$,
      C.\,Fedeli$^{4,5,6}$, S. Matarrese$^{7,8}$, L.\,Moscardini$^{4,5,6}$,
\\~\\
\LARGE{\rm P.\,Rosati$^{9}$ and J.\,Weller$^{10,11}$}\\~\\
$^1$ Dipartimento di Fisica, Sezione di Astronomia, Universit\`a di
Trieste, Via Tiepolo 11, I-34143 Trieste, Italy\\ 
$^2$ INAF-Osservatorio Astronomico di Trieste, Via Tiepolo 11, I-34143
Trieste, Italy\\ 
$^3$ INFN, Sezione di Trieste, Via Valerio 2, I-34127 Trieste, Italy\\
$^4$ Dipartimento di Astronomia, Universit\`a di Bologna, Via Ranzani
1, I-40127 Bologna, Italy\\
$^5$ INAF-Osservatorio Astronomico di Bologna, Via Ranzani 1, I-40127
Bologna, Italy\\
$^6$ INFN, Sezione di Bologna, Viale Berti Pichat 6/2, I-40127
Bologna, Italy\\
$^7$ Dipartimento di Fisica ``G. Galilei'', Universit\`a di Padova, via Marzolo
8, I-35131, Padova, Italy\\
$^8$ INFN -- Sezione di Padova, via Marzolo 8, I-35131, Padova, Italy\\
$^9$ ESO-European Southern Observatory, D-85748
Garching bei M\"unchen, Germany\\
$^{10}$ University Observatory, Ludwig-Maximillians University Munich,
Scheinerstr. 1, 81679 Munich, Germany\\
$^{11}$ Excellence Cluster Universe, Boltzmannstr. 2, 85748 Garching, Germany
}

\maketitle

\begin{abstract}
We present forecasts for constraints on deviations from Gaussian
  distribution of primordial density perturbations from future
  high--sensitivity X--ray surveys of galaxy clusters. Our analysis is
  based on computing the Fisher--Matrix for number counts and
  large-scale power spectrum of clusters.  The surveys that we
  consider have high--sensitivity and wide--area to detect about
  $2.5\times 10^5$ extended sources, and to provide reliable
  measurements of robust mass proxies for about $2\times 10^4$
  clusters. Based on the so-called self-calibration approach, and
  including Planck priors in our analysis, we constrain at once nine
  cosmological parameters and four nuisance parameters, which define
  the relation between cluster mass and X--ray flux. Because of the
  scale dependence of large--scale bias induced by local--shape
  non--Gaussianity, we find that the power spectrum provides strong
  constraints on the non--Gaussianity $\fnl$ parameter, which
  complement the stringent constraints on the power spectrum
  normalization, $\sigma_8$, from the number counts. To quantify the
  joint constraints on the two parameters, $\sigma_8$ and $\fnl$, that
  specify the timing of structure formation for a fixed background
  expansion, we define the figure-of-merit $\fom=\left(
  \det\left[Cov(\sigma_8,\fnl)\right]\right)^{-1/2}$. We find that our
  surveys constrain deviations from Gaussianity with a precision of
  $\Delta \fnl\simeq 10$ at $1\sigma$ confidence level, with $\fom
  \simeq 39$. We point out that constraints on $\fnl$ are weakly
  sensitive to the uncertainties in the knowledge of the nuisance
  parameters. As an application of non--Gaussian constraints from
  available data, we analyse the impact of positive skewness on the
  occurrence of XMMU-J2235, a massive distant cluster recently
  discovered at $z\simeq 1.4$.  We confirm that in a WMAP-7 Gaussian
  $\Lambda$CDM cosmology, within the survey volume, $\simeq 5\times
  10^{-3}$ objects like this are expected to be found. To increase the
  probability of finding such a cluster by a factor of at least 10,
  one needs to evade either the available constraints on $\fnl$ or on
  the power spectrum normalization $\sigma_8$.

\end{abstract}

\section{Introduction}
The standard inflationary scenario, based on the single scalar field
slow-roll paradigm, predicts primordial density perturbations to be
virtually indistinguishable from a Gaussian distribution. However, a
number of variants of inflation have been proposed which are able to
generate a certain amount of non-Gaussianity
\citep[e.g.,][]{bartolo04,Chen10}. Therefore,
testing to what precision we can measure possible deviations from
Gaussianity with available and future observations has important
implications on our understanding of the mechanism that seeded density
fluctuations in the early Universe. Analyses of the Cosmic Microwave
Background (CMB) provide at present the tighest constraints on the
amount of allowed non-Gaussianity.  A number of analyses based on the
WMAP data converge to indicate consistency with the Gaussian
assumption (e.g., \citealt{komatsu10} and references therein; cf. also
\citealt{Yadav_Wandelt07}). While data from the Planck satellite are
expected to further tighten such constraints
\citep[e.g.,][]{Yadav_etal07,Liguori_etal10}, it is worth
understanding whether non-Gaussianity can be probed by large--scale
structure observations \citep[e.g.,][]{SL08.1,verde10}.\\

Non-Gaussian perturbations are expected to leave their imprint also on the
pattern of structure growth at least in two different ways.  First, we
expect that a positively skewed distribution provides an enhanced
probability of finding large overdensities. This translates into an
enhanced probability of forming large collapsed structures at high
redshift, thereby changing the timing of structure formation and the
shape and evolution of the mass function of dark-matter halos.

After the first pioneering studies of the effect of non-Gaussiantiy on
the mass function
\citep[e.g.][]{Matarrese_etal86,Colafrancesco_etal89,Borgani_etal90},
a number of analyses have been carried out since the beginning of
2000s \citep[e.g.,][]{matarrese00,mathis04,kang07,sefusatti07,grossi07,
maggiore09}. Furthermore, the
realization of large-scale cosmological simulations with non-Gaussian
initial conditions have recently provided a validation of the
non-Gaussian correction to be applied to the Gaussian mass function
\citep[e.g.,][]{grossi07,
DA08.1,DE08.1,grossi09,Giannatonio_Porciani09,pillepich10}.
More recently non-Gaussianity effects on the large-scale distribution of
collapsed halos were studied:
it has been demostrated
\citep[e.g.,][]{DA08.1,matarrese08,valageas09,Lam_Sheth09} that
non-Gaussianity affects the large-scale clustering of halos in such a
way that the linear biasing parameter acquires a scale dependence. This
modifies, in a detectable way, the power spectrum of the distribution of any
tracer of cosmic structures at small wavenumbers and
offers a unique way of testing the nature of primordial fluctuations.

The evolution of the mass function of galaxy clusters identified in
X--ray surveys has been extensively used in the past
to constrain cosmological models
\citep[e.g.,][]{borgani01,rosati02,schuecker03,voit05}. These studies
have recently attracted renewed interest, thanks to detailed follow-up
observations of clusters selected from ROSAT observations, they have
been carried out either to constrain the Dark Energy equation of state
within the Gaussian paradigm \citep[e.g.,][]{vikhlinin09c,mantz09I},
or to test possible deviations from standard gravity
\citep[e.g.,][]{Schmidt_etal09,rapetti09}. \citet{jimenez09} have
recently analyzed the effect of non-Gaussianity on the population of
massive high-redshift clusters, like the one recently discovered by
\citet{jee09} at $z\simeq 1.4$. It is important to remember that the
cosmological constraints obtained from clusters so far have been
derived from small ROSAT--based samples, containing $\sim 100$
clusters at $z < 1$. It is therefore easy to imagine the vast margin
of improvements offered by next generation X-ray surveys which will
detect $\sim 10^5$ clusters to $z \sim 2$.

These surveys should cover a large enough volume at high redshift to
test non-Gaussianity in the regimes where its effects are clearer,
namely the high-mass tail of the mass function and the large-scale
power spectrum of the cluster distribution. \citet{fedeli09} and
\citet{roncarelli10} presented predictions for the number counts and
clustering of galaxy clusters expected from the eROSITA X--ray survey
and from the Sunyaev--Zeldovich SPT survey. While these analyses
confirmed the potential of these surveys to provide interesting
constraints on non-Gaussian models, they did not include detailed
forecasts on the constraints on non-Gaussian models and a detailed
assessment of the effect of uncertainties in the scaling relations
between cluster masses and observables. \citet{oguri09} followed the
self--calibration approach by \citet{lima05} (see also
\citealt{majumdar04,battey03}) to forecast the capability of future
optical cluster surveys to constrain non--Gaussian models. This study
showed that combining number counts and clustering of galaxy clusters
can potentially provide quite strong constraints on deviations from
Gaussianity.

The aim of this paper is to derive forecasts, based on the
Fisher--Matrix approach, on the capability of future X--ray cluster
surveys to constrain deviations from Gaussian perturbations. Besides
focusing on the characteristics of X-ray, rather than optically
selected samples, our analysis differs from that by  \citet{oguri09}
for the method to include information from large--scale
clustering. \citet{oguri09} adopted the approach by
\citet{lima05} where clustering is included by accounting for
fluctuations of cluster counts within cells having a fixed angular
size. This implies that, at each redshift, clustering information is
restricted to one physical scale. In our analysis, we follow the
approach originally presented by \citet{tegmark97} (see also
\citealt{feldmann94,majumdar03}), in which the clustering
Fisher-Matrix is computed for the allowed range of wavenumbers,
by weighting them according to the effective volume covered by the
surveys. 

Another distinctive aspect of our analysis is that it is based on
X-ray surveys of next generation, whose sensitivity and angular
resolution are high enough to warrant an accurate measurement of
robust mass proxies, related to the cluster gas mass and X--ray
temperature for a large number of clusters. As we will discuss in the
following, surveys with these characteristics can be provided by an
already proposed X--ray telescope, which joins a large collecting area
to a large field-of-view and a high angular resolution over the entire
field of view \citep[e.g.,][]{giacconi09,vikhlinin09b}. The great
advantage of having a similar survey is that there is no need to
assume any external follow-up observation for a subset of identified
clusters. Moreover, the possibility to define a flux-limit down to
which measuring accurate mass proxies for all clusters allows one to
set robust priors on the scaling relations between cluster mass and
observables, which is one of the main source of uncertainty in the
cosmological application of galaxy clusters
\citep[e.g.,][]{albrecht09}.

In principle, the method used in our analysis can be applied
  to any cluster surveys, including optical and SZ ones. Although so
  far X-ray surveys have been mostly used for cosmological
  applications of clusters, upcoming large optical and SZ surveys
  promise to provide an important contribution to this field. Our method only requires
  a well defined selection function and calibrated mass proxies. Since
  cluster surveys at different wavelengths have different efficiencies
  probing different mass ranges at different redshifts, they will
  ultimately provide complementary approaches to the derivation of
  cosmological constraints.

The plan of this paper is as follows. In Section 2 we summarize the
formalism to compute non--Gaussian corrections to the mass function
and the linear bias parameter of collapsed halos. In Section 3 we
describe our approach to compute the Fisher Matrix for both the number
counts and the power spectrum of galaxy clusters. In Section 4 we first
describe how we compute the selection function and the redshift
distribution expected for the X--ray surveys, then we present
the results in terms of constraints on the parameter
space defined by the non--Gaussianity parameter, $\fnl$, and the
power spectrum normalizations, $\sigma_8$. Section 5 is devoted to the
discussion of these results. In this section we will also discuss the
competing effects of non-Gaussianity and normalization of the power
spectrum on the expected number of clusters at $z>1.4$, which have a mass of, at
least, $5\times
10^{14}M_\odot$, as the one recently studied by \citet{jee09}. We
summarize our main conclusions in Section 6.

\section{Non-Gaussian initial conditions}
\label{s:nong}

Generalizations of the most standard model of inflation \citep{guth81}
give rise to seed primordial density fluctuations that follow a
non-Gaussian probability distribution \citep[e.g.,][ for
reviews]{bartolo04,Chen10}. A particularly convenient way to
parametrize the deviation of this distribution from the Gaussian one 
consists of writing the Bardeen's gauge invariant potential $\Phi$ as
the sum of a linear Gaussian term and a quadratic correction
\citep{salopek90,gangui94,verde00,komatsu01},

\begin{equation}\label{eqn:ng}
\Phi = \Phi_\mathrm{G} + f_\mathrm{NL}*\left( \Phi_\mathrm{G}^2 - \langle
\Phi_\mathrm{G}^2 \rangle \right).
\end{equation}
In Eq. (\ref{eqn:ng}), the symbol $*$ denotes convolution between
functions  and reduces to simple multiplication only in the particular
case of constant $f_\mathrm{NL}$, while in general it is a
function of the scale. Note that on scales smaller than the Hubble
radius the function $\Phi$ equals minus the Newtonian peculiar
gravitational potential. The fundamental parameter $f_\mathrm{NL}$
denotes the amplitude of the deviation from Gaussianity, and is
related to the skewness of the distribution (see below).

We adopt, in the following, the Large Scale Structure convention (as
opposed to the CMB convention, see
\citealt{afshordi08,pillepich10,carbone08,grossi09}) for defining the
parameter $f_\mathrm{NL}$. This means, among other things, that the
constraints given on $f_\mathrm{NL}$ by the CMB have to be raised,
according to the linear growth of structures, as $f_\mathrm{NL} =
g(+\infty) f_\mathrm{NL}^\mathrm{CMB}/g(0) \simeq 1.3
f_\mathrm{NL}^\mathrm{CMB}$, where $g(z)$ is the linear growth
suppression factor with respect to the Einstein-de Sitter cosmology.

If the distribution of primordial density (and potential)
perturbations is not Gaussian, it cannot be fully described by a power
spectrum expressed as $P_\Phi({\bf k}) = Bk^{n-4}$ (where $k = \|{\bf
  k}\|$), but we need higher-order moments such as the bispectrum
$B_\mathrm{\Phi}({\bf k}_1,{\bf k}_2,{\bf k}_3)$. In particular,
different models of inflation give rise to different shapes of the
bispectrum. In the following we adopt one particular
shape, called \emph{local shape}, means that the bispectrum is maximized
for configurations in which one of the three momenta is much smaller
than the other two ("squeezed" configurations). Inflationary models
exist that produce different shapes for the primordial bispectrum,
e.g., the \emph{equilateral shape} (see \citealt{crociani09,fedeli09}
for applications), or the \emph{enfolded shape}
\citep{holman08,meerburg09,verde09}, however, the local shape is the
one giving the largest effects especially on bias \citep{fedeli09,taruya08},
hence we limit our analysis to this case only.

In this case the parameter $f_\mathrm{NL}$ is a dimensionless constant
and the bispectrum can be written as \citep[e.g.,][]{creminelli07,loverde08}
\begin{equation}
B_\Phi({\bf k}_1,{\bf k}_2,{\bf k}_3) = 2f_\mathrm{NL} B^2 \left[
k_1^{n-4}k_2^{n-4} + k_1^{n-4}k_3^{n-4} + k_2^{n-4}k_3^{n-4} \right],
\end{equation}
where $n$ is the primordial spectral index and $B$ is the amplitude of
the spectrum $P_\Phi(\bf{k})$, related to the amplitude $A$ of the
power spectrum of density fluctuations, $P(k)=Ak^n$, by the relation
\begin{equation}
B = \frac{9}{4} A H_0^4\Omega_{\mathrm{m},0}^2.
\end{equation}

Two of the ingredients that critically affect the observed properties
of the cluster population are influenced by non-Gaussian initial
conditions: the halo mass function and the linear bias of dark-matter
halos.

\subsection{Mass function}
\label{s:mf}

For the mass function of cosmic structures in non-Gaussian cosmologies
different prescriptions exist. The one adopted here is taken from
\citet{loverde08}, where the authors approximated the probability
distribution for the smoothed dark-matter density field using the
Edgeworth expansion and then performed the integral of the probability
distribution for threshold crossing exactly on the first few terms of
the expansion itself. 
The result for the \citet{press74} (PS) mass function
reads
\begin{eqnarray}\label{eqn:mfps}
n_\mathrm{NG}(M,z) &=& - \sqrt{\frac{2}{\pi}} \frac{\bar{\rho}}{M} \exp\left[
-\frac{\delta_\mathrm{c}^2(z)}{2\sigma_M^2} \right] \left[ \frac{d\ln
\sigma_M}{dM} \left( \frac{\delta_\mathrm{c}(z)}{\sigma_M} + \right.\right.
\nonumber\\
&+& \left. \left. \frac{S_3\sigma_M}{6} \left(
\frac{\delta_\mathrm{c}^4(z)}{\sigma^4_M}
-2\frac{\delta^2_\mathrm{c}(z)}{\sigma^2_M} -1\right) \right) + \right.
\nonumber\\
&+& \left. \frac{1}{6} \frac{dS_3}{dM}\sigma_M \left(
\frac{\delta^2_\mathrm{c}(z)}{\sigma^2_M} -1\right) \right].
\end{eqnarray}
In Eq. (\ref{eqn:mfps}) $\delta_\mathrm{c}(z) \equiv
\Delta_\mathrm{c}/D(z)$, where $D(z)$ is the linear growth factor,
$\sigma_M$ is the \emph{rms} of primordial density fluctuations on the
scale corresponding to mass $M$, while $S_3(M) \equiv f_{\mathrm{NL}}
\mu_3(M)/\sigma_M^4$ is the normalized skewness. The third-order
moment $\mu_3(M)$ can be written as

\begin{equation}
\mu_3(M) = \int_{\mathbb{R}^9} \mathcal{M}_R(k_1) \mathcal{M}_R(k_2)
\mathcal{M}_R(k_3) B_\Phi({\bf k}_1,{\bf k}_2,{\bf k}_3) \frac{d{\bf
    k}_1d{\bf k}_2d{\bf k}_3}{(2\pi)^9}\,. 
\end{equation}
The function $\mathcal{M}_R(k)$ relates the Fourier transform of
density fluctuations smoothed on some scale $R$ to the relative
peculiar potential, and is defined as
\begin{equation}
\mathcal{M}_R(k) \equiv \frac{2}{3} \frac{T(k)k^2}{H_0^2
  \Omega_{\mathrm{m},0}} W_R(k)\,, 
\end{equation}
where $T(k)$ is, in our analysis, the matter transfer function \citep{eisenstein98}  and $W_R(k)$
is the
top-hat window function. 
Eq.\ref{eqn:mfps} gives the correction to the PS mass
  function. However, we know that the PS mass function provides only
  an approximate fit to the results of N-body simulations. The commonly
  adopted procedure is then to assume that the same correction
  appearing in Eq.(\ref{eqn:mfps}) can be applied to the best-fit
  Gaussian mass function to derive an accurate expression for the
  non-Gaussian one:
\begin{equation}
\label{eq:MF_fin}
n(M,z) = n_\mathrm{(G)}(M,z) \frac{n_\mathrm{NG}(M,z)}{n_\mathrm{PS}(M,z)}\,.
\end{equation}
Here $n_{\mathrm{(G)}}(M,z)$ is the mass function in the reference
Gaussian model computed according to the \citet{sheth02} recipe, while
$n_\mathrm{NG}(M,z)$ and $n_\mathrm{PS}(M,z)$ represent the
\citet{press74} mass functions in the non-Gaussian
(Eq. \ref{eqn:mfps}) and reference Gaussian model respectively.

As already discussed, different prescriptions can be found in the
literature that give different expressions for the mass function in
non-Gaussian models. For instance, \citet{matarrese00} used the saddle
point approximation to compute the probability distribution of
threshold crossing, and then truncated the resulting expression to the
skewness. The resulting mass function is however in agreement with the
one obtained by \citet{loverde08}, as are the other recipes found in
the literature.

\citet{grossi09} have shown that these analytic expressions are in
agreement with N-body cosmological simulations, provided the linear
threshold for collapse is corrected for ellipsoidal density
perturbations, according to $\Delta_\mathrm{c} \rightarrow
\Delta_\mathrm{c} \sqrt{q}$, with $q=0.75$. We adopted this correction
in our calculations.

\subsection{Bias}
\label{s:bias}

The halo bias acquires an extra scale dependence due to primordial
non-Gaussianity, that can be written as \citep{matarrese08}

\begin{equation}
b(M,z,k) = b^\mathrm{(G)}(M,z) + \Delta b(M,z,k),
\end{equation}
where 

\begin{equation}
\Delta b(M,z,k) = \left[ b^\mathrm{(G)}(M,z)-1 \right]
\delta_\mathrm{c}(z) \Gamma_R(k)\,. 
\end{equation}
The term $\Gamma_R(k)$ encapsulates the dependence on the scale and on
the mass, and can be written for a local bispectrum shape as
\begin{eqnarray}
\Gamma_R(k) &=&
\frac{2f_\mathrm{NL}}{8\pi^2\mathcal{M}_R(k)\sigma_R^2}
\int_0^{+\infty}d\zeta\,\zeta^2\mathcal{M}_R(\zeta)
P_\Phi(\zeta)\times 
\nonumber\\
&\times& \left\{ \int_{-1}^1
  d\mu\,\mathcal{M}_R\left(\sqrt{\alpha}\right)
  \left[\frac{P_\Phi\left(\sqrt{\alpha}\right)}{P_\Phi(k)}+2\right]
\right\},
\end{eqnarray}
where  $\alpha = \zeta^2 + k^2 + 2\mu\zeta k$. 
The linear bias in the reference Gaussian model can be written as 

\begin{eqnarray}
b^\mathrm{(G)}(M,z) &=& 1+\frac{a\Delta_\mathrm{c}}{D^2(z)\sigma_M^2} -
\frac{1}{\Delta_\mathrm{c}}+
\nonumber\\
&+&\frac{2p}{\Delta_\mathrm{c}} \left[
\frac{\left(D(z)\sigma_M\right)^{2p}}{\left(D(z)\sigma_M\right)^{2p}+\left(\sqrt
{a}\Delta_\mathrm{c} \right)^{2p}} \right],
\label{eq:bias}
\end{eqnarray}

according to the prescriptions of \citet{mo96,sheth99,sheth01}, and
with parameters set to $p = 0.3$ and $a = 0.75$.

In order to obtain agreement with the results of numerical simulations
it is necessary to correct the linear overdensity for collapse, this
time according to $\Delta_\mathrm{c} \rightarrow \Delta_\mathrm{c} q$ with
$q=0.75$ 
\citep{grossi09}.  Semi-analytic results, with this correction, are also
in agreement with the numerical results of \citet{pillepich10} (see also
\citealt{DE08.1}). We adopted this correction in the remainder of our
calculations. For figures showing the scale, mass and redshift
dependence of this correction to the linear bias we refer to
\citet{taruya08} and \citet{fedeli09}.

\section{Fisher matrix analysis}
\label{s:fm}
The Fisher Matrix formalism can be used to understand how accurately we
can estimate the values of a vector of parameters $\textbf{p}$ for a
given model from one or more data sets, under the assumption that all
parameters follow a Gaussian distribution \citep[e.g.,][]{Cash79}.

The information Fisher Matrix (FM hereafter) is defined as
\begin{equation}
F_{\alpha \beta}\equiv - \left\langle \frac{ \partial^2 \ln {\cal
      L}}{\partial p_{\alpha}\partial p _{\beta}} \right\rangle ,
\label{eq:FM}
\end{equation}
where ${\cal L}$ is the likelihood of an observable (in our case the
number of galaxy clusters in a given redshift and mass range or the
averaged power spectrum of the cluster distribution).


\subsection{Number counts}
\label{s:fm_nc}
Following the approach of \citet{holder01} and \citet{majumdar03},
the Fisher matrix for the number of clusters, $N_{l,m}$, within the
$l$-th redshift bin and $m$-th bin in observed mass $M^{ob}$, can be
written as
\begin{equation}
  F^N_{\alpha \beta}= \sum_{l,m} \frac{\partial N_{l,m}}{\partial
    p_\alpha}\frac{\partial N_{l,m}}{\partial p_\beta}
  \frac{1}{N_{l,m}}\,, 
\label{eq:fm_nc}
\end{equation}
where the sums over $l$ and $m$ run over redshift and mass intervals,
respectively. With this notation, it is $M^{ob}_{l,m=0}=M_{\rm thr}(z)$,
where $M_{\rm thr}(z)$ is defined as the threshold value of the
observed mass for a cluster to be included in the survey. Due to the
selection function of any X--ray flux-limited survey, the value of
$M_{\rm thr}(z)$ depends on redshift (see Section \ref{s:sur}). Therefore, the
number of mass bins and their extent in our
analysis will change with redshift accordingly.

We write the number of clusters expected in a survey having a sky
coverage $\Delta\Omega$ with observed mass between $M^{ob}_{l,m}$ and
$M^{ob}_{l,m+1}$ and redshift between $z_l$ and $z_{l+1}$ as
\begin{eqnarray}
N_{l,m} & = & \Delta\Omega \int_{z_l}^{z_{l+1}}dz\,{dV\over dz d\Omega}
\int_{M^{ob}_{l,m}}^{M^{ob}_{l,m+1}}dM^{ob} \nonumber \\ 
& &\int_0^\infty dM \,n(M,z)\,p(M^{ob}\|M)\,.
\label{eq:nln}
\end{eqnarray}
In the above equation $dV/dz$ is the cosmology--dependent comoving volume
element per unity redshift interval and solid angle, $n(M,z)$ the
mass function, i.e. the number density of clusters with true mass $M$ at
redshift $z$ (see Section \ref{s:mf}). 

As already mentioned, we assume in the following the expression by
\cite{sheth99} for the Gaussian halo mass function. We remind here that other
calibrations of the halo mass function from simulations have been
presented by several authors
\citep[e.g.,][]{jenkins01,warren06,tinker08,crocce09}. While using the
best-calibrated mass function is in fact important when deriving
cosmological constraints from real data \citep[e.g.,][]{wu09}, it has
only a minor impact when deriving forecasts on cosmological
constraints. Indeed, what matters for the latter is the total number
of clusters expected in a given cosmological model, which is far more
sensitive to the choice of reference cosmological and nuisance
parameters than to the details of the mass function fitting function.

Following \cite{lima05} we assign to each cluster with true mass
$M$ a probability $p(M^{ob}\|M)$ of having an observed mass $M^{ob}$, as
inferred from a given mass proxy. Under the assumption of a log-normal
distribution for the intrinsic scatter in the relation between true
and observed mass, with
variance $\sigma^2_{\ln M}$, the expression for the probability is
\begin{equation}
p(M^{ob}\|M)\,=\,{\exp[-x^2(M^{ob})]\over \sqrt{\left( 2\pi \sigma^2_{\ln
M}\right) }}\,,
\label{eq:prob}
\end{equation}
where
\begin{equation}
x(M^{ob})\, =\, \frac{\ln M^{ob}-B_M-\ln M}{\sqrt{\left( 2 \sigma^2_{\ln
M}\right) }}\,.
\label{eq:m_mo}
\end{equation}
Here we allow the relation between true and observed mass to be
characterized not only by an intrinsic scatter, but also by a
systematic bias in the mass estimate, whose fractional value is given
by $B_M$. By inserting Eq.(\ref{eq:prob}) into Eq.(\ref{eq:nln}) for
the cluster counts in a given mass and redshift interval, we obtain
\begin{eqnarray}
N_{l,m} & = & {\Delta\Omega \over 2}\int_{z_l}^{z_{l+1}}dz\,{dV\over
  dzd\Omega} 
\int_0^\infty dM\, n(M,z)\nonumber \\
& \times & \left[{\rm erfc}(x_m)-{\rm erfc}(x_{m+1}) \right]\,
\label{eq:nln2}
\end{eqnarray}
with $x_m=x(M^{ob}_{l,m})$ and erfc$(x)$ the complementary error
function. We note that the possibility of factorising the sky-coverage
outside the integration relies on the assumption that clusters in our
survey are detected over the same area of the sky down to the survey
completeness limit. The above expression can be easily generalized to
include the possibility of a ``flux''--dependent sky coverage.

We remark that we neglect clustering contribution to the noise
(i.e. cosmic variance). Indeed, the Wide and the Medium surveys
(see Table \ref{t:sur}) covers large enough area over which
cosmic variance is negligible. This may not be case for the Deep
survey, which however, as we shall see, does not bring much
cosmological constraints (see Section \ref{s:sur}).

\subsection{Power spectrum}
\label{fm_pk}
In order to include in our analysis the information from the
clustering of galaxy clusters, we follow the approach by
\citet{majumdar04}.  We define the Fisher Matrix for the power spectrum
of galaxy clusters as
\begin{equation}
F_{\alpha \beta}={1\over (2\pi)^2}\,\sum_{l,m} 
{\partial \ln{\bar{P}_{cl}(k_m,z_l)}\over \partial p_\alpha} 
{\partial \ln{\bar{P}_{cl}(k_m,z_l)}\over \partial p_\beta}\,
V^{eff}_{l,m}k_m^2\Delta k,
\label{eq:fm_pk}
\end{equation}
where the sums in $l$ and $m$ run over redshift and wavenumber $k$ bins,
respectively \citep{tegmark97,feldmann94}. 
In the above equations $\bar{P}_{cl}$ is the average cluster power
spectrum calculated within the given redshift interval,
\begin{equation}
\bar{P}^{cl}_{l,m}(k,z_i) = \frac{\int^{z_{l+1}}_{z_l} dz
  \,\frac{dV}{dz} \,N^2(z)\, P^{cl}(k,z)}{\int^{z_{l+1}}_{z_l} dz
  \,\frac{dV}{dz}\, N^2(z) }\,.
\label{eq:barpk}
\end{equation}
This amounts to weight the cluster power spectrum, $P^{cl}(k,z)$,
according to the square of the number density of clusters, $N(z)$,
that are included in the survey at redshift $z$.  In turn, the cluster
power spectrum $P^{cl}(k,z)$ is expressed in terms of power spectrum,
$P(k,z)$, of the cosmic density fluctuations according to
$P^{cl}(k,z)=b_{eff}^2(k,z)\,P(k,z)$; here the effective bias is
defined as the linear bias, introduced in Sect. 2, weighted by the
mass function,
\begin{equation}
  b_{eff}(z,k) = \frac{\int_0^{\infty} dM n(M,z)\,{\rm erfc}[x(M_{\rm thr})]\,
b(M,z,k)} {\int_0^{\infty} dM\,n(M,z)\,{\rm erfc}[x(M_{\rm thr})]}\,.
\label{eq:beff}
\end{equation}
The bias parameter $b(M,z,k)$ acquires the dependence on the
wavenumber $k$ predicted by non--Gaussian models (see
Sect. \ref{s:bias}). 
Finally, the quantity $V^{eff}(k,z)$ in Eq.(\ref{eq:fm_pk}) is the
effective volume accessible by the survey at redshift $z$ at
wavenumber $k$. This effective volume is weighted by the shot noise
level $1/N(z)$, so that
\begin{equation}
V^{eff}(k,z)\,=\,V_0(z)\,\left[
\frac{N(z)\bar{P}_{cl}(k,z)}{1+N(z)\bar{P}_{cl}(k,z)}\right]^2 \, , 
\label{eq:veff}
\end{equation}
with $V_0(z)$ the total comoving volume covered by the redshift bin
centred on $z$. In this way, constraints at redshift $z$ are mostly
contributed by wavemodes $k$, which maximize $N(z)\bar{P}_{cl}(k,z)$
and make $V^{eff}$ approach $V_0$.

An alternative approach to include clustering information in deriving
FM survey forecasts has been proposed by \citet{lima05} and applied
also by \citet{oguri09} for constraints on non--Gaussian models from
cluster surveys. In this approach, one makes a partition of the sky
area covered by a survey into regular cells of a fixed angular size
and then computes the fluctuations in the cluster counts within such
cells.  Since this method does not explicitly include the covariance
between counts within different cells, it only samples clustering at a
fixed angular scale (i.e. at a single physical scale for a fixed
redshift). On the other hand, extracting the full information content
in the scale dependence of the power spectrum is quite important when
constraining non-Gaussian models, whose unique signature is given by
the scale--dependent bias. \citet{cunha10} used the
  count-in-cell approach by also including the information from the
  covariance. Therefore, the information
  on the large-scale power spectrum, in the count-in-cell approach, is conveyed by the covariance
  terms. In our approach the different scales are weighted by the
  effective volume, defined by Eq.(\ref{eq:veff}).

In our analysis we assume the following reference values for the
cosmological parameters, consistent with the WMAP-7 best--fitting
model \citep{komatsu10}: $\Omega_m=0.28$ for the present-day matter
density parameter, $\sigma_8=0.81$ for the normalization of the power
spectrum, $\Omega_k=0$ for the contribution from the curvature,
$w(a)=w_0+(1-a) \; w_a$ with $w_0=-0.99$ and $w_a=0$ for the Dark
Energy equation of state, $\Omega_{\rm b}=0.046$ for the contribution
of baryons to the density parameter, $h=0.70$ for the Hubble
parameter, $n=0.96$ for the primordial spectral index and $\fnl=0$ for
the non--Gaussianity parameter.  Therefore, we have in total 9
cosmological parameters, which are left free to vary in the
computation of the number counts and power spectrum Fisher Matrices of
Eqs.(\ref{eq:fm_nc}) and (\ref{eq:fm_pk}).

In the following, all the results presented are based on adding
the Fisher Matrix for the Planck CMB experiment to those from the
cluster surveys. We derive the cosmological constraints from Planck
following the description laid out by the DETF \cite{albrecht09} and use
the method described in \cite{rassat08}. We conservatively assume that
we will only use the 143 GHz channel as science channel. This channel
has a beam of $\theta_{\rm fwhm}=7.1'$ and sensitivities of $\sigma_T
= 2.2 \mu K/K$ and $\sigma_P = 4.2\mu K/K$.  We take
$f_{\rm sky} = 0.80$ as the sky fraction in order to account for
galactic foregrounds. We use as a minimum $\ell$-mode, $\ell_{\rm
  min}=30$ in order to avoid problems with polarization
foregrounds. As described in the DETF report \citep{albrecht09} we
choose as fiducial parameter set $\vec{\theta}= (\omega_m,
\theta_S,\ln A_S, \omega_b, n_S, \tau)$, where $\theta_S$ is the
angular size of the sound horizon at last scattering, $\ln A_S$ is the
logarithm of the primordial amplitude of scalar perturbations and
$\tau$ is the optical depth due to reionization. After marginalization
over the optical depth we then calculate the Planck CMB Fisher matrix
in the parameters $(\Omega_m, \Omega_{de},h, \sigma_8, \Omega_b, w_0,
w_a, n_S)$ by using the appropriate Jacobian of the involved parameter
transformation \citep{rassat08}. We point out that the Planck FM is
computed for Gaussian perturbations. Therefore, while it adds quite
strong constraints on the other cosmological parameters, especially on
the curvature, it does not add any constraints on $\fnl$. 

\section{Analysis and results}
\label{s:res}
\subsection{Characteristics of the surveys}
\label{s:sur}
The commonly adopted procedure to estimate cosmological forecasts for
X--ray surveys is based on calibrating a flux limit for cluster
detection (which generally corresponds to the detection of few tens of
net photon counts in an extended source), and to use fluxes as proxies
to cluster masses. In order to account for the uncertain knowledge of
the relation between X--ray luminosity and mass, several authors have
proposed to follow the so--called self--calibration method
\citep[e.g.,][]{majumdar03,majumdar04,lima04,lima05}. In this approach
the relation between mass and observable is defined up to an intrinsic
scatter, also including the possibility of a systematic bias in the
estimate of cluster masses. The parameters defining the relation
between mass and observable are then treated as fitting ``nuisance''
parameters to be determined along with the relevant cosmological
parameters.

An alternative approach, adopted to derive cosmological constraints
from cluster surveys, is, instead, based on restricting the analysis to a
relatively small subset of galaxy clusters for which deeper X--ray
follow-up observations provide measures of mass proxies which are more
closely related to cluster masses
\citep[e.g.,][]{vikhlinin09a,mantz09I}. Examples of such robust mass
proxies are the so-called $Y_X=M_{\rm gas}T_X$, defined as the product
of gas mass and temperature at a given radius
\citep[e.g.,][]{kravtsov06}, or $M_{\rm gas}$ itself. They are robust
in the sense that, based on simulations,
their relation with cluster mass has an intrinsic
scatter of 10 per cent or less, especially when emission from cluster
cores can be resolved and/or removed. While having obvious advantages,
the limitation of relying on such low-scatter mass proxies is that
they can be measured for a relatively small number of clusters. In
this sense, one has to compromise between the requirement to keep
under control the systematics in mass measurements and the need to
cover a relatively large redshift baseline with adequate statistics.

Ideally, one should carry out an X--ray survey at such a good
sensitivity that low--scatter mass proxies can be measured in survey
mode for a large number of galaxy clusters. This would allow
an ``educated'' self--calibration analysis
of the cluster surveys, in which the systematics in mass measurements
can be kept under control.

In the following, we derive forecasts for three X--ray surveys, which
are complementary in terms of sensitivity and sky coverage, inspired
to the survey strategy devised for the Wide Field X--ray Telescope
(WFXT\footnote{http://www.eso.org/$\sim$prosati/WFXT/Overview.html,
  http://wfxt.pha.jhu.edu/}), recently proposed to the Astro-2010
Decadal Survey panel
\citep[e.g.,][]{murray10,giacconi09,vikhlinin09b}. This telescope
combines a large collecting area and field of view with a sharp PSF over
the entire field of view. Thanks to these characteristics, the WFXT
has the potential of detecting a large number of clusters out to
$z\sim 2$ and of measuring $Y_X$ or $M_{\rm gas}$ with good precision
down to relatively low fluxes. While describing the characteristics of
the WFXT is outside the scope of this paper, we point out that such an
instrument would have the capability of measuring low-scatter mass
proxies down 
to cluster fluxes which are comparable to fluxes at which
clusters are just detected in current and future X--ray
telescopes.
Based on the results presented by \citet{giacconi09}, we
give in Table \ref{t:sur} the limiting fluxes (in the 0.5--2 keV
energy band) at which WFXT will detect a cluster as an extended
source in three surveys: a Wide Survey, which
will cover all the extragalactic sky (20000 deg$^2$) with a sensitivity
$\sim 500$ times better than the ROSAT All Sky Survey
\citep[e.g.,][]{RASS}; a Medium Survey, which will reach over 3000
deg$^2$ flux limits comparable to those of the deep Chandra and XMM
deep COSMOS fields \citep[][]{cappelluti_etal09}; a Deep Survey that
will reach over 100 deg$^2$ a sensitivity similar to those of the
deepest Chandra pointings. Thanks to the large collecting area and
field of view of WFXT, and taking advantage of its good angular
resolution (5 arcsec half energy width, approximately constant over the whole
field-of-view), these surveys could be completed within a five--year
mission duration \citep{murray10}.

\begin{table} 
\centering
\caption{Characteristics of the X-ray surveys (see also
  \protect\citealt{giacconi09}. Column 2: sky coverage
  $\Omega$ (in sq.deg.); Column 3: flux limits for detection of
  extended sources in the [0.5-2] keV energy band (units of
  $10^{-14}$erg s$^{-1}$cm$^{-2}$); Column 4: flux limits defining
  the {\em bright} subsamples (see text).} 
\begin{tabular}{lccc}
 & $\Omega$ & $F_{\rm det}$ & $F_{\rm br}$ \\ 
\hline 
Wide  & 20000 & 0.5 & 15.0\\
Medium& 3000  & 0.1 & 3.0 \\
Deep  & 100    & 0.01& 0.3 \\
\end{tabular}
\label{t:sur}
\end{table}

In principle, a unique flux limit is not sufficient to define a
completeness criterion in an X--ray survey. In fact, due to vignetting
and PSF variation
 with off-axis angle, the flux limit for the
detection of a source at a given signal-to-noise varies across the
field of view. For this reason, rather than a flux limit, one should
calibrate a flux-dependent sky coverage. Owing the approximate uniformity of the WFXT PSF,
 we expect such a sky coverage to be
quite steep around the flux limits reported in Table \ref{t:sur}, so
that we ignore its flux--dependence in the following analysis.  In
order to convert these flux limits into mass limits, we use the
relation between X--ray luminosity and $M_{500}$ calibrated by
\citet{maughan07}, where masses are recovered from $Y_X$, using
Chandra data for 115 clusters in the redshift range $0.1<z<1.3$. Among
the fitting expressions reported in Table 1 of that paper, we choose
the relation between $L_X$ and $M_{500}$, obtained
without excising the core region within $0.15 \, R_{500}$:
\begin{equation}
 L_X \, = \, C\, E(z) \left( \frac {M_{500}}{4 \times 10^{14}\,M_{\odot}}
\right)^B\,,
\end{equation}
with $C=5.6$, $B=1.96$.

The reason for
this choice is that we did not attempt to model the core contribution
in computing the flux limits reported in Table \ref{t:sur}.

\begin{figure}
\hspace{-0.5truecm}
\includegraphics[width=0.45\textwidth,angle=270]{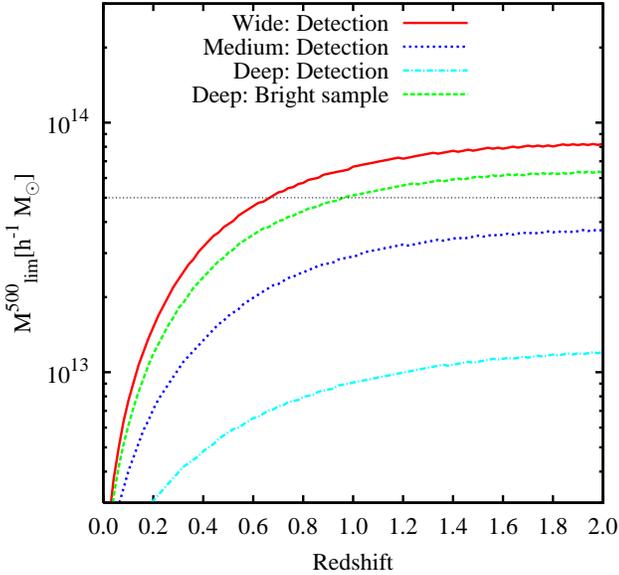}
\caption{The redshift dependence of the $M_{500}$ mass thresholds,
  corresponding to the flux limits for cluster detection for the three
  surveys, as reported in Table 1: the Wide, Medium and Deep surveys
  are shown with the solid (red), the dotted (blue) and dot-dashed
  (cyan) curves, respectively. The short-dashed (green) curve
  corresponds to a flux-limit which is 30 times brighter than the flux
  limit for cluster detection in the Deep survey. The horizontal
  dotted line marks the mass limit of $M_{500}=5\times 10^{13}\msun$
  below which we discard clusters in our analysis.}
\label{fig:sf}
\end{figure}

We show in Figure \ref{fig:sf} the redshift dependence of the limiting
mass\footnote{Here and in the following we indicate with
  $M_{\Delta}$ the mass contained within the radius $R_\Delta$,
  encompassing an average density of $\Delta$ times the critical
  cosmic density $\rho_{cr}$.} $M_{500}$ associated to the survey flux
limits. The value of the virial mass, which is the relevant quantity
entering in the mass function choosed in our analysis, is obtained
from $M_{500}$ , following \citet {hu03}, by adopting the NFW halo
density profile \citep{navarro97} for the reference cosmological
model, assuming $c=5$ for the concentration parameter \citep[see also
][]{shang09}. Also shown in Figure \ref{fig:sf} with the green
short-dashed curve is the mass limit corresponding to a flux 30 times
brighter than the flux limit for cluster detection in the Deep
Survey. Since cluster detection in the Deep survey corresponds to
about 200 net counts, such a mass limit is for clusters for which
about 6000 counts would be available. With such a large number of
counts one has a precise measurement of robust mass proxies such as
$Y_X$ and $M_{gas}$. We point out that clusters identified in the
Deep Survey allows one to calibrate mass proxies down to fluxes which
are lower than the flux for cluster identification in the Wide
Survey. Such brighter flux limits define the {\em bright samples} of
clusters, for which a direct measurement of a mass proxy can be
carried out within the same surveys.  Extrapolating the
  $M$--$L_X$ relation by \cite{maughan07} at faint fluxes of our
  surveys would imply unrealistically small mass limits at low
  redshift. For this reason, we decided to use a lower limit of
  $M_{500}=5\times 10^{13}\msun$ in the definition of the selection
  function (shown with the horizontal dotted line in
  Figure \ref{fig:sf}). In fact, this is comparable to the lowest mass
  down to which mass proxies have been calibrated so far
  \citep[e.g.,][]{vikhlinin09a}. We also point out that, for the sake
  of simplicity, we do not include the cosmology dependence of the
  selection function in our analysis.

\subsection{Nuisance parameters}
\label{s:nuis}
Besides the nine cosmological parameters, our FM analysis should also
constrain the nuisance parameters which specify the redshift
dependence of the fraction mass bias $B_M$ and the intrinsic scatter
$\sigma_{\ln M}$
(in our analysis we do not consider the case of a possible mass
dependence of these parameters). According to \cite{lima05}, we assume
the following parametrization for such redshift dependencies:
\begin{eqnarray}
B_M(z) & = & B_{M,0}(1+z)^\alpha \nonumber \\
\sigma_{\ln M}(z) & = & \sigma_{\ln M,0}(1+z)^\beta \,.
\label{eq:nuis}
\end{eqnarray}
In this way, we have four nuisance parameters, $B_{M,0}$, $\sigma_{\ln
  M,0}$, $\alpha$ and $\beta$. A negative value for $B_M$ corresponds
to a mass understimate and, therefore, to a smaller number of clusters
included in a survey, for a fixed selection function.  The presence of
the mass bias accounts for the possibility of a violation of
hydrostatic equilibrium in the estimate of X--ray masses, on which the
observable--mass scaling relation is calibrated. A number of
independent analyses of a variety of cosmological hydrodynamic
simulations of galaxy clusters converge to indicate that hydrostatic
mass estimators provide underestimates of true mass within $R_{500}$
by about 10--15 per cent
\citep[e.g.,][]{rasia05,nagai07,ameglio09,piffaretti08}. Quite
reassuringly, such results also agree with the observational results
on the comparison between cluster masses estimated with weak lensing
and with X-ray data \citep[e.g.,][]{mahdavi08}. In the following we
assume $B_{M,0}=-0.15$ as a reference value for the mass bias and
regarding its evolution, we take $\alpha=0$ as a reference value. As
for the intrinsic scatter, it has the effect of increasing the number
of clusters included in the survey. In fact, the number of low-mass
clusters that are up-scattered above the survey mass limit is always
larger than the number of rarer high-mass clusters which are
down-scattered below the same mass limit \citep[e.g.,][ and references
therein]{Cunha09}. As a reference value, we assume $\sigma_{\ln
  M,0}=0.25$, consistent with the instrinsic scatter in the
$M_{500}$--$L_X$ relation measured by \cite{maughan07}, with $\beta=0$
for its evolution. We stress here that, following
  \citet{lima05}, we use the variance $\sigma_{lnM}^2$ and not the
  scatter as the parameter to be varied in our Fisher matrix
  analisys. In fact, this quantities controls the excess of
  up-scattered and down-scattered clusters with respect the total
  number.

In the following we will not assume any prior for these four nuisance
parameters in our reference analysis. We will refer to it as the {\em no
prior} analysis. 

On the other hand, already available data allow one to set
constraints on the value of the mass bias from the comparison between
X--ray and lensing cluster mass measurements.  For instance,
\citet{vikhlinin09a} compared weak lensing and Chandra X--ray mass
measurements for a rather small sample of low--$z$ clusters and
concluded that the mass scale can already be calibrated with a
statistical uncertainty of about 10 per cent. A similar result has
been obtained by \citet{zhang10} from the comparison of
XMM--Newton X--ray masses and weak lensing masses for a set of 12
nearby clusters. Owing to these results, we also consider the case of
the mass bias parameter to be known with a precision $\Delta
B_{M,0}=0.05$.  An improvement by only a factor of two with respect to
the present in the calibration of the cluster mass scale is probably
overconservative, owing to the orders-of-magnitude increase in the
number of clusters with precise X--ray mass measurements to be
provided by the three surveys and precise lensing mass measurements
from both ground-based and space telescopes. As for the evolution, we
assume $\Delta \alpha=1$ as a prior, which would correspond to an
uncertainty in the mass bias calibration at $z=1$ comparable to that
calibrated at present for nearby clusters. Regarding the prior on the
intrinsic scatter, we assume $\Delta \sigma_{\ln M,0}=0.1$ and
$\Delta\beta=1$. We expect these to be rather conservative
choices, in view of the large number of clusters that should be made
available by future X--ray and optical/near-IR surveys at both low and
high redshift. In the following, we refer to these choices of the
priors for the nuisance parameters as the {\em weak prior} analysis.

Finally, we also refer to the {\em strong prior} analysis for the
case in which nuisance parameters are assumed to be known with little
uncertainty and are thus kept fixed to their reference values. While
this assumption is expected to be unrealistic for fluxes reaching the
limiting flux for cluster detection in the three surveys, it may be
rather plausible in case we restrict the analysis to the {\em bright}
subsamples.

While we use the {\em no prior} choice as a reference for the analysis
of the surveys down to the detection limit, we will show how much we
would gain in terms of constraining power by using instead the {\em
  weak prior} and the {\em strong prior} assumptions. Finally, we will
present results for the {\em bright} surveys for the case of {\em
  strong priors} on the nuisance parameters. This will allow us to
judge the trade-off between having statistically richer (i.e. lower
flux limit) surveys with a less controlled calibration of the cluster
mass-luminosity conversion, and a brighter flux limit with better
controlled nuisance parameters.\\

\begin{figure}
\includegraphics[width=0.45\textwidth,angle=270]{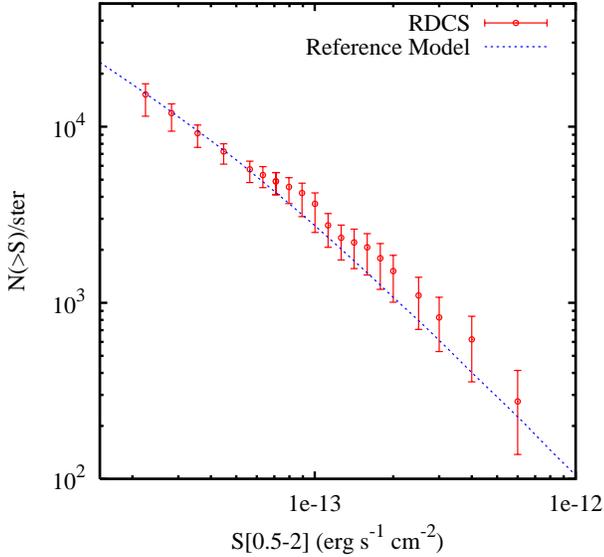}
\caption{The comparison between observed cumulative cluster flux
  number counts (symbols with errorbars) and predictions from the
  reference model (dotted curve, see text). Observational results
  refer to counts in the [0.5-2] keV band from the ROSAT Deep Cluster
  Survey \protect\citep{rosati02}, with errorbars corresponding to
  $1\sigma$ Poissonian uncertainties.}
\label{fig:ns}
\end{figure}

As a first check, we verify that our reference model provides a good
fit to present data. We show in Figure \ref{fig:ns} the comparison
between the cluster flux number counts observed in the [0.5-2] keV
energy band from the ROSAT Deep Cluster Survey
\citep{rosati98,rosati02} and the prediction of our reference
model. This is not surprising, owing to the fact that the reference
cosmological parameters agree with constraints based on the evolution
of the cluster mass function
\citep[e.g.,][]{borgani01,vikhlinin09c,mantz09I}. The
good agreement between available observational data and predictions of
our reference model indicates that the latter can be used to provide a
realistic extrapolation of the evolution of the cluster mass function
over redshift and mass ranges which are not probed by currently
available data.

\begin{figure*}
\hbox{
\includegraphics[width=0.45\textwidth,angle=270]{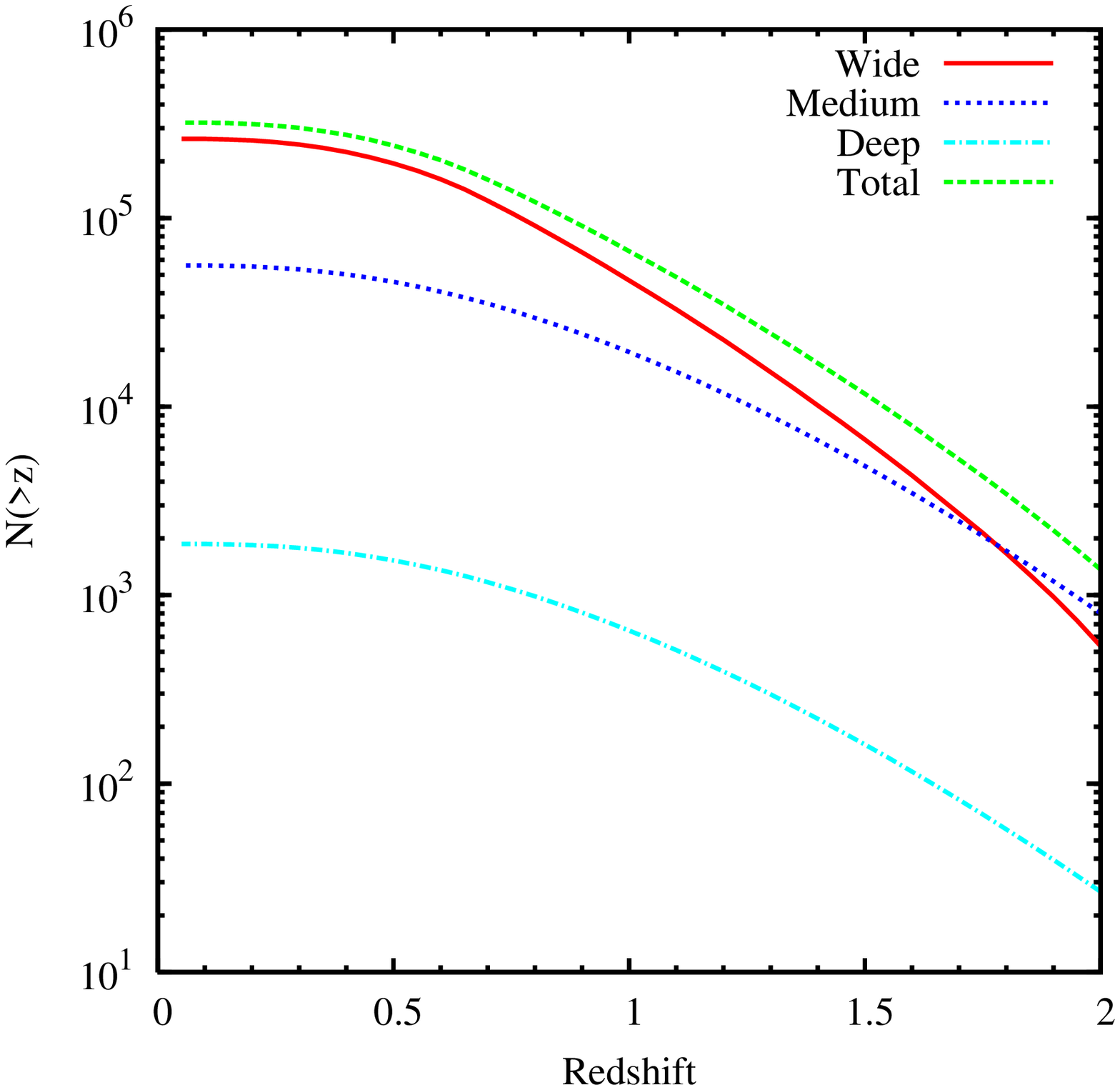}
\hspace{-3.truecm}
\includegraphics[width=0.45\textwidth,angle=270]{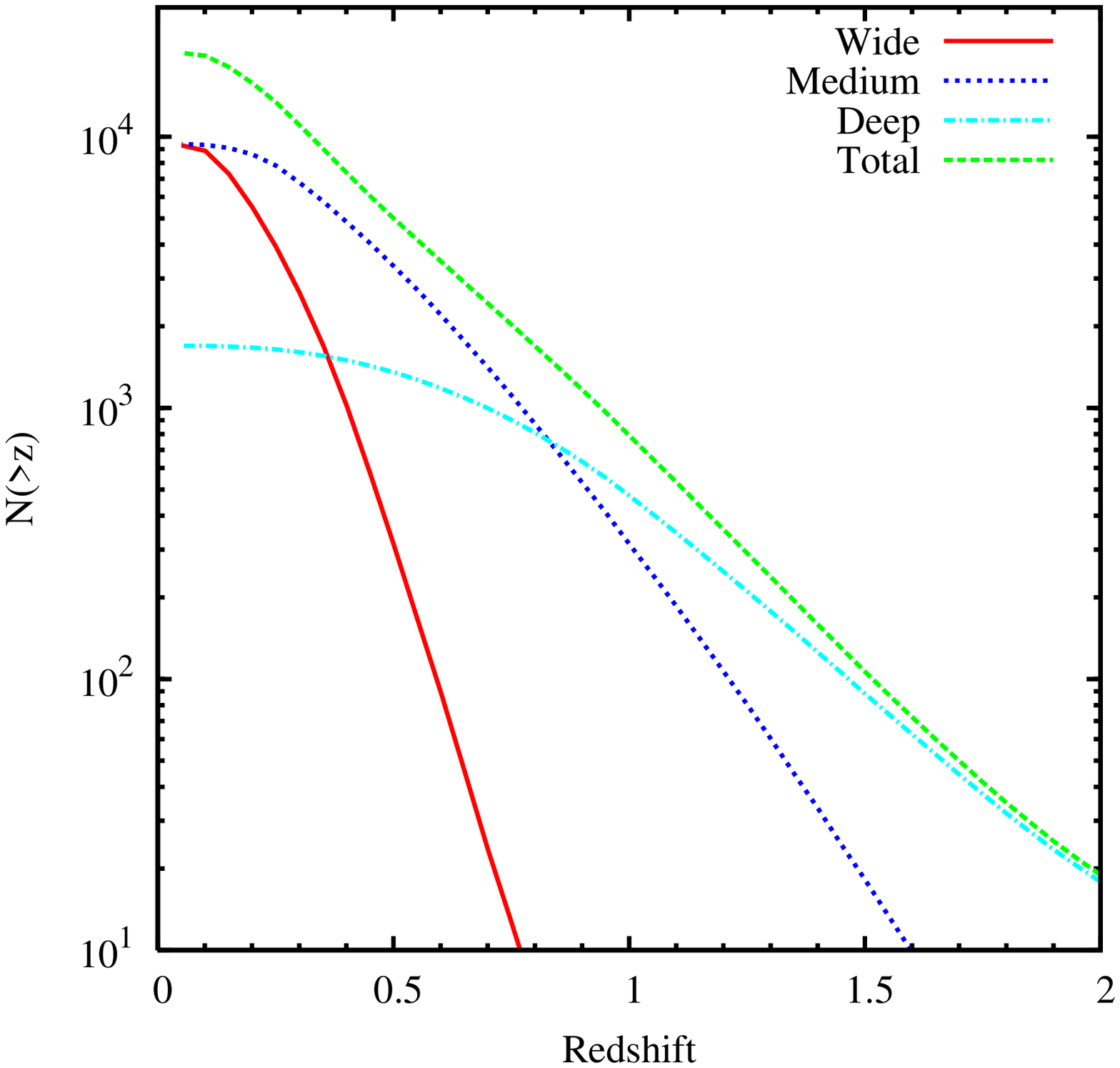}
}
\caption{The cumulative redshift distribution for the three
  surveys. The left panel is for all clusters to be detected down to
  the survey flux limits, while the right panel is for the clusters in
  the {\em bright samples}, corresponding to a 30 times higher flux
  limit. In both panels solid (red), dotted (blue) and dot-dashed
  (cyan) curves represet the Wide, Medium and Deep surveys,
  respectively, while the short-dashed (green) curve represents the sum of
  the three.}
\label{fig:nz}
\end{figure*}

We show in Figure \ref{fig:nz} the cumulative redshift distributions
for the clusters to be detected in the three surveys (left panel) and
for the {\em bright samples} (right panel). Overall, the three WFXT
surveys would yield about $3\times 10^6$ detected clusters, out of
which $\sim 7.5\times 10^4$ clusters should be found at $z>1$. This
will provide an improvement by about four orders or magnitude with
respect to the $\sim 10$ $z>1$ clusters currently confirmed. At the
same time, we expect to have about $2\times 10^4$ clusters with
robustly measured mass proxies, of which $\sim 4000$ would lie at
$z>0.5$. This would increase by more than two orders of magnitude the
number of clusters for which mass proxies have been measured above
this redshift, after intensive follow-up Chandra observations of
clusters identified in ROSAT--based surveys
\citep[e.g.,][]{vikhlinin09a,mantz09II,ettori09}. We stress that,
despite the small area covered, the Deep survey provides the dominant
contribution to the {\em bright sample} at $z>1$. This highlights the
important role that the Deep survey has in providing mass proxies at
high redshift.  Although predicting the number of extremely distant
clusters is highly uncertain, owing to the unknown evolution of the
mass luminosity relation above $z\simeq 1$, we foresee that $\sim
10^3$ clusters would be detected at $z\magcir 2$, with mass
measurements available for few tens of them.  In order to quantify the
increase in sensitivity provided by the WFXT surveys with respect to
currently planned X--ray missions, we computed the redshift
distributions expected for the surveys to be carried out by the
eROSITA
satellite\footnote{http://www.mpe.mpg.de/erosita/MDD-6.pdf}. In the
three years of operation eROSITA is expected to find, for our
reference model, $\simeq 4000$ clusters at $z>1$, while virtually no
clusters with precise measurements of mass proxies would be found at
$z>0.5$ in the {\em bright surveys} having flux limits 30 times higher
than for detection.

An important lesson learned from the Chandra follow-up observations of
distant clusters is the fundamental relevance of a sharp PSF to excise
the contribution of cool cores in the measurements of mass proxies.
Indeed, as shown by a number of authors
\citep[e.g.,][]{maughan07,pratt09}, excising the core contribution
suppresses by a substantial amount the intrinsic scatter in the
relation between cluster X--ray observables and masses. The sharp PSF
with which the WFXT would carry out the surveys, should guarantee a
good control of the cool core contribution in the calibration of mass
proxies, even to $z \sim 1.5$ without the need of follow-up
observations with other higher resolution X--ray telescopes.
We also emphasize that a sharp PSF, such that expected for
  WFXT, has the additional benefit of easing the subtraction
  of the point source contribution when measuring cluster
  fluxes. This allows an accurate assessment of
  the survey flux completeness.

\subsection{Constraints on non-Gaussianity}
Having defined the reference cosmological model and mass--flux
conversion, we present now forecasts on constraints from non-Gaussian
models. These results will be shown in terms of constraints on the
$\sigma_8$--$\fnl$ plane after marginalizing over the other
cosmological and nuisance parameters. The reason for this choice is
that, for a fixed Friedmann background, $\sigma_8$ and $\fnl$ are the
two parameters which determine the timing of structure formation and,
therefore, the evolution of number density and large-scale clustering
of galaxy clusters. In the following, we will always show constraints
on the $\sigma_8$--$\fnl$ plane at the 68 per cent confidence level.

In analogy with the figure-of-merit introduced to quantify the
constraining power of an experiment for the Dark Energy equation of
state \citep[e.g.,][]{albrecht06,albrecht09,wang08}, we
introduce a figure-of-merit for the timing of structure formation in
non-Gaussian models:
\be
\fom\,=\,\left(
  \det\left[Cov(\sigma_8,\fnl)\right]\right)^{-1/2}\,,
\label{eq:fom}
\ee where $Cov(\sigma_8,\fnl)$ is the covariance matrix between
$\sigma_8$ and $\fnl$, which is obtained by inverting the FM and
marginalizing over all the other parameters.

In the computation of the FM of Eq.(\ref{eq:fm_nc}), the $N_{l,n}$
number counts are computed out to $z=2$ within $60$ constant redshift
bins and within observed mass bins having width $\Delta \log M= 0.1$, extending
from the lowest mass limit determined by the selection function (see
Fig. \ref{fig:sf}) and an upper mass limit of $10^{16}h^{-1}M_\odot$.
We verified that a finer binning does not add more
  information in the Fisher Matrix. As for the redshift bins, their
  size is larger than the redshift errors obtainable from optical
  spectrscopy, and comparable to the typical uncertainties
  from red-sequence redshift estimates \citep{gladders07}.  

As for the computation of the FM for the power spectrum of
Eq.(\ref{eq:fm_pk}), we used wavenumbers in the range $k_{max}\ge k\ge
0.001\,\MpcI$, independent of redshift. Using an arbitrary small value
of $k$ does not change the final results, since extremely large wave
modes are not sampled by the surveys and, therefore, do not provide
any contribution to the FM. As for the value of $k_{max}$, the chosen
value should represent a compromise between the needs of maximising
the amount of information to be extracted from the survey and of
avoiding the contribution of small--scale modes where the validity of
the linear bias model is compromised by the onset non-linearity
\citep{percival09}. In the following, we will assume
$k_{max}=0.3\,\MpcI$, and will also show the sensitivity of the
results to this choice.

As for the redshift binning, the average cluster power spectrum
defined by Eq.(\ref{eq:barpk}) is computed by integrating over
redshift intervals having constant width $\Delta z=0.2$.  This coarser
binning, with respect to that used for the analysis of number counts,
is dictated by a compromise between the need of extracting the maximum
amount of information from the clustering evolution and the request of
negligible covariance between adjacent $z$-intervals
\citep[e.g.,][]{stril09}. Indeed, the contribution from different
$z$-bins can be added in the defining of FM of Eq.(\ref{eq:fm_pk})
only if they carry statistically independent information.

\begin{figure}
\includegraphics[width=0.47\textwidth]{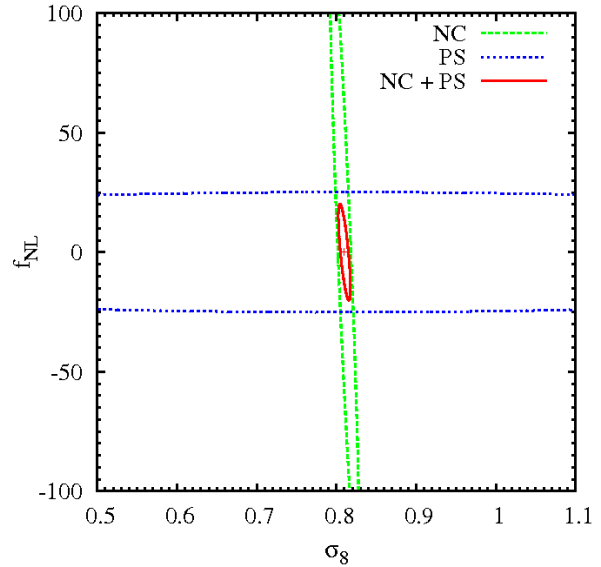}
\caption{Constraints at the 68 per cent confidence level on
  non-Gaussian parameter $\fnl$ and power spectrum normalization
  $\sigma_{8}$ coming from number counts alone (short-dashed green
  curve), power spectrum alone (dotted blue curve) and from the
  combination of the two (solid red curve). The analysis refer to the
  clusters detected in the Wide Survey. We marginalized over all
  the other parameters. {\em No prior} is assumed for the
  values of the nuisance parameters. The Fisher Matrix from Planck
  experiment is included in the calculation of all constraints.}
\label{fig:wide}
\end{figure}

We show in Figure \ref{fig:wide} the constraints on the $\fnl$ and
$\sigma_8$ parameters computed from the number counts and from the
power spectrum within the Wide survey, by assuming {\em no prior} on
the values of the nuisance parameters. This plot clearly demonstrates
the strong complementarity that number counts and large-scale
clustering have to constrain $\sigma_8$ and $\fnl$: while number
counts are highly sensitive to the value of $\sigma_8$, the weak
sensitivity of the high-end of the mass function to non-Gaussianity
\citep[e.g.,][and references therein]{fedeli09} provides only very
weak constraints on $\fnl$; conversely, the scale-dependence of bias
makes the power spectrum a powerful diagnostic for non--Gaussianity,
while providing only loose constraints on $\sigma_8$.

\begin{figure}
\hspace{-0.2truecm}
\includegraphics[width=0.47\textwidth]{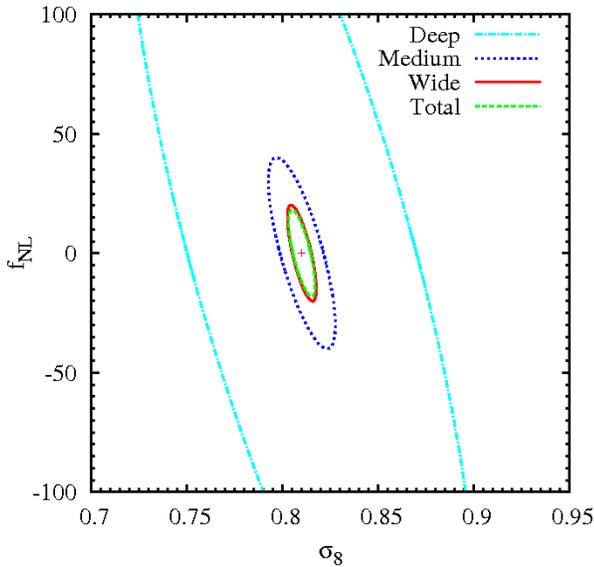}
\caption{Constraints at the 68 per cent confidence level on non--Gaussian
parameter $\fnl$ and power
  spectrum normalization $\sigma_8$ from the Deep, Medium and Wide
  surveys (dot-dashed
  cyan, dotted blue and solid red curves,
  respectively), by combining number counts and power spectrum
  information, by using no priors on the nuisance parameters.  Also
  shown with the short-dashed green curve are the constraints obtained
  from the combination of the three surveys. {\em No prior} is assumed for the
  values of the nuisance parameters. The Fisher Matrix from Planck
  experiment is included in the calculation of all constraints.}
\label{fig:det}
\end{figure}

If we combine all the information obtainable from the three surveys,
we obtain the constraints shown in Figure \ref{fig:det}. Most of the
constraining power is provided by the Wide survey, with only little
information on structure formation timing carried by the Medium and
Deep surveys. There are two main reasons for this. First, the Wide
survey provides the largest statistical baseline out to $z\simeq 1.5$,
when including all clusters down to the mass limit corresponding to
detection (see left panel of Figure \ref{fig:nz}). This implies a better
determined mass function and, therefore, stronger constraints on
$\sigma_8$. Second, the larger area coverage of the Wide survey
allows it to better sample long-wavelength modes, where the
scale--dependence of the bias induced by non--Gaussianity can be
better assessed, thus turning into stronger constraints on
$\fnl$. As shown in Table \ref{t:fom} the value of $\fom$ for the
combination of the three surveys is in fact dominated by the Wide
Survey.

\begin{table} 
\centering
\caption{Figure-of-merit of structure formation timing, $\fom$
  [see Eq. \protect\ref{eq:fom}], and r.m.s. uncertainty in the
  non-Gaussian parameter, $\sigma_{\fnl}$, for the three surveys,
  and for their combination, assuming different priors for the
  nuisance parameters. Columns 3--6 show the results for the Wide,
  Medium and Deep Surveys, and for the combination of the three.}
\begin{tabular}{llcccc}
			&         & Wide 	& Medium 	& Deep 	& Total \\
\hline \\
Detection - no pr.	& $\fom$ 	& 33.1 & 8.5    & 0.4  	& 39.2  \\
			& $\sigma_{\fnl}$&11.3 & 18.5   & 84.2  & 10.4  \\
Detection - weak pr.	& $\fom$ 	& 33.3 & 8.8    & 0.6  & 39.4  \\
			& $\sigma_{\fnl}$&11.3 & 18.4   & 84.3  & 10.4  \\
Detection - strong pr.	& $\fom$ 	& 157.2 & 49.3    & 3.0  & 183.2  \\
			& $\sigma_{\fnl}$&11.2 & 18.0   & 80.9  & 10.3  \\
Bright - strong pr. 	& $\fom$ 	&  7.3 & 14.3    & 3.0  & 22.3  \\
			& $\sigma_{\fnl}$&55.9 & 45.9   & 85.7  & 33.8  \\
\hline
\end{tabular}
\label{t:fom}
\end{table}

\begin{figure}
\includegraphics[width=0.47\textwidth]{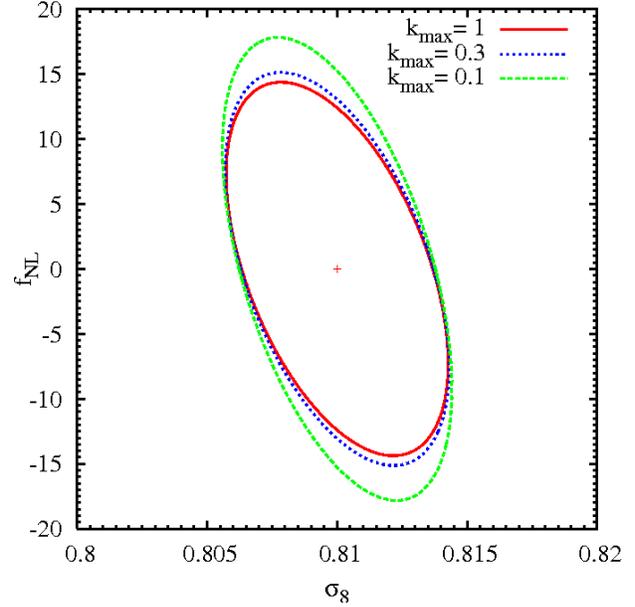}
\caption{Constraints at the 68 per cent confidence level on non-Gaussian
parameter $f_{nl}$ and power
  spectrum normalization $\sigma_8$ from the combination of the three
  surveys, when changing the maximum value of the wavenumber,
  $k_{max}$, for which power spectrum information are included in the
  Fisher Matrix of Eq.(\protect\ref{eq:fm_pk}). Solid (red), dotted
  (blue) and short-dashed (green) curves correspond to $k_{max}=1$, 0.3
  and 0.1$\MpcI$, respectively. {\em No prior} is assumed for the
  values of the nuisance parameters. The Fisher Matrix from Planck
  experiment is included in the calculation of all constraints.}
\label{fig:km}
\end{figure}

As already mentioned, we have assumed in our analysis $k_{max} = 0.3\MpcI$. In
order to quantify the sensitivity of our
results to the adopted $k_{max}$ value, we show in Figure \ref{fig:km}
how the constraints change for $k_{max}=0.1\MpcI$ and
$k_{max}=1\MpcI$. The smaller value is close to the scale of
non--linearity at $z=0$, although it is probably too conservative at
high redshift, $z\sim 1$. As expected, decreasing $k_{max}$ makes the
constraints slightly looser, due to the lower amount of information
included in the Fisher Matrix of Eq.(\ref{eq:fm_pk}). Correspondingly,
the value of the figure-of-merit decreases from $\fom=39.2$ to $33.1$,
with the uncertainty on $\fnl$ increasing only from
$\sigma_{\fnl}=10.4$ to $12.1$. Increasing instead $k_{max}$ to $1\MpcI$
does not lead to any significant improvement of the constraints. In
fact, given the level of Poisson noise associated to the cluster
distribution, high frequency modes are not adequately sampled and,
therefore, adding them to the analysis does not add significant
information.

\begin{figure*}
\hbox{
\hspace{-0.3truecm}
\includegraphics[width=0.46\textwidth,angle=270]{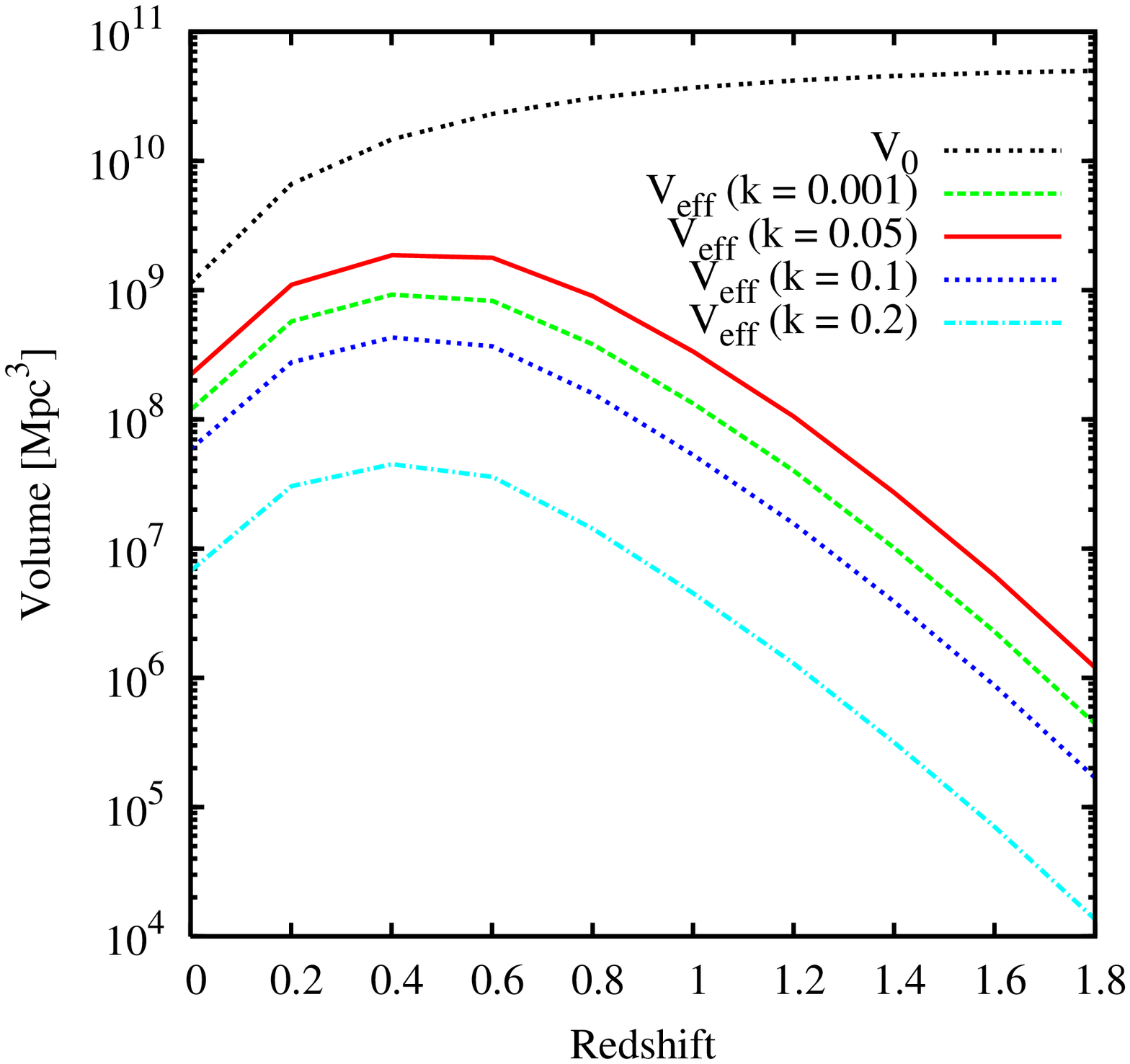}
\hspace{-3.truecm}
\includegraphics[width=0.46\textwidth,angle=270]{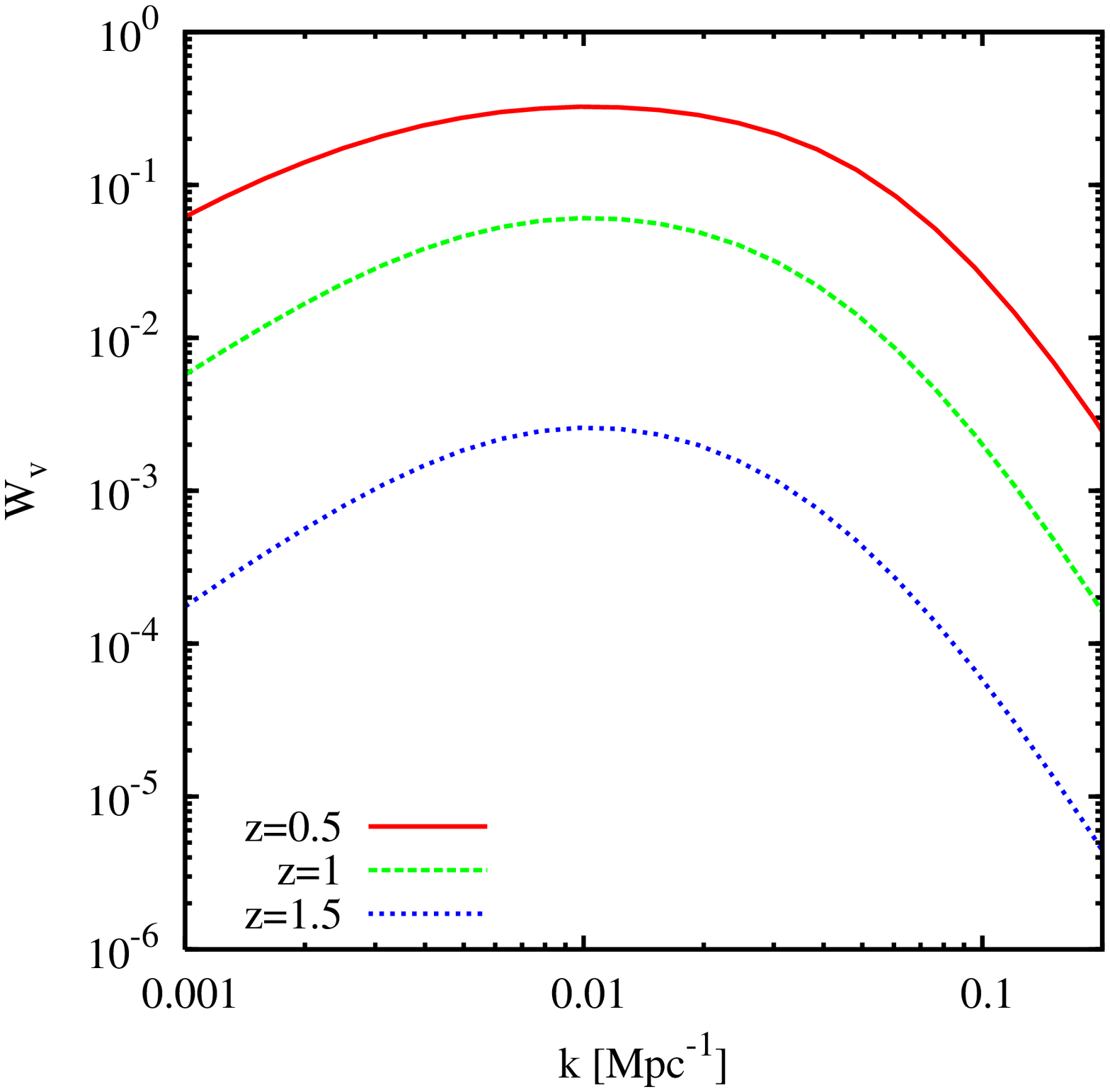}
}
\caption{Left panel: the redshift dependence of the effective volume,
  defined as in Eq.(\protect\ref{eq:veff}), within redshift intervals
  of constant width $\Delta z=0.2$, for four values of the wavenumber
  $k$. Short-dashed (green), solid (red), dotted (blue) and
  dot-dashed (cyan) curves correspond to $k=0.001$, 0.05, 0.1 and
  0.2$\MpcI$, respectively. Right panel: the dependence on the
  wavenumber of the weight $W_v(k,z)$, defined as in
  Eq.(\protect\ref{eq:ww}), at three different redshifts. Solid (red),
  dashed (green) and dotted (blue) curves are for $z=0.5$, 1 and 1,5,
  respectively.}
\label{fig:pesiz}
\end{figure*}

The contribution of information to the Fisher Matrix carried by the
power spectrum at different redshifts and wavenumbers can be
understood by looking at the dependence of the effective volume,
$V_{eff}$, on the power spectrum, which is set by the bias parameter,
and on the level of Poisson noise, which is set by the number density
of clusters. Following Eq.(\ref{eq:veff}), we define the quantity 
\begin{equation}
W_v(k,z)= \left[ \frac{N(z)
    \bar{P}_{cl}(k,z)}{1+N(z)\bar{P}_{cl}(k,z)}\right]^2\,,
\label{eq:ww}
\end{equation}
which gives the weight carried by the wavenumber $k$ to the
computation of the clustering Fisher Matrix at redshift $z$. In the
left panel of Figure \ref{fig:pesiz} we show the redshift dependence
of the effective volume computed within redshift intervals of constant
width $\Delta z=0.2$, for different values of $k$, and compare them to
the total comoving volume computed within the same redshift
intervals. The effective volume lies always well
below the total comoving volume: this is the consequence of the
relatively low value of the cluster number density, which makes
Poisson noise always dominating.  While the total comoving volume
$V_0$ increases with redshift, the effective volume $V_{eff}$ starts
declining after reaching a maximum, at $z\simeq 0.5$, for all
wavenumbers. As for the dependence on $k$, at a fixed redshift, the
value of $V_{eff}$ decreases for both very high and very low values of
$k$. As shown in the right panel of Figure \ref{fig:pesiz}, the value of
the weight function $W_v(k,z)$ is maximized at $k\simeq 0.01\MpcI$. In
fact, for $W_V \ll 1$ (i.e. $N(z)\bar{P}_{cl}(k,z)\ll 1$), the
$k$--dependence of $W_V$ reflects that of $\bar{P}_{cl}$.
Poisson noise is, again, responsible for the low values of $W_V$, well below
unity.  Decreasing of the level of this noise would require increasing
the number density of objects to be included in the survey.  This
could be accomplished in principle by decreasing the mass
threshold. However, this would require bringing into the surveys
low--mass clusters and groups, for which our parametrization of the
mass--observable relation may not still be valid.

\begin{figure}
\includegraphics[width=0.47\textwidth]{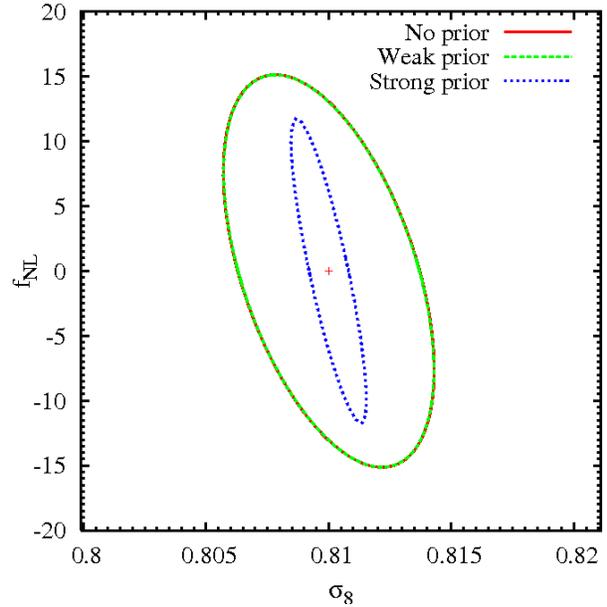}
\caption{Constraints at the 68 per cent confidence level on the
  non-Gaussian parameter $\fnl$ and power spectrum normalization
  $\sigma_8$ by assuming {\em no prior} (solid red curve), {\em weak
    prior} (dashed green) and {\em strong prior} (dotted blue) on the
  nuisance parameters. All constraints are obtained by combining
  cluster number counts and power spectrum information for the three
  surveys together. The Fisher Matrix from Planck
  experiment is included in the calculation of all constraints.}
\label{fig:prior}
\end{figure}

So far, we presented results by assuming prior on cosmological parameters from
Planck experiment and {\em no prior} knowledge on
the nuisance parameters. We want to stress that, as already discussed in
Sect. \ref{s:nuis}, this is probably too much a conservative approach,
in view of the calibration of the relation between robust mass proxies
(e.g., $Y_X$ and $M_{gas}$) and X--ray luminosity for a large number
of clusters within the planned surveys. In Figure \ref{fig:prior} we
show the effect of assuming a prior knowledge of the nuisance
parameters. If we assume the {\em weak priors} for these parameters
(see Sect. \ref{s:nuis}), constraints are only slightly
improved. Quite interestingly, even assuming the {\em strong prior}
(i.e. nuisance parameters fixed) improves the constraints on
$\sigma_8$, while having a smaller impact on those for $\fnl$. Indeed,
we find that the error on non-Gaussianity only decrease from
$\sigma_{\fnl}\simeq 10.4$ to 10.3 when passing from the {\em no
  prior} to the {\em strong prior} assumption, while the
figure-of-merit increases from $\fom\simeq 39.2$ to
$\simeq 183.2$ (table: \ref{t:fom}).

\begin{figure}
\includegraphics[width=0.35\textwidth,angle=270]{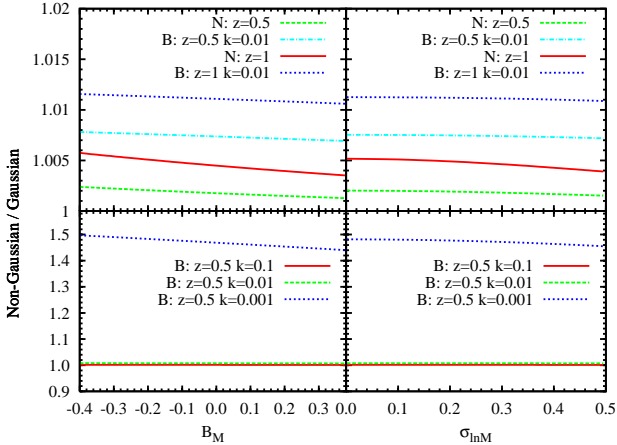}
\caption{Effect of changing the values of the nuisance parameters of
  mass bias $B_M$ (left panels) and intrinsic mass scatter
  $\sigma_{\ln M}$ on the deviations of number counts and effective
  bias from the Gaussian case. The results shown here are obtained by
  assuming a mass-limit of $10^{14}M_\odot$. The reference value of
  non--Gaussianity assumed here is $\fnl=10$. N is the ratio between
  the redshift distributions in the Gaussian and non-Gaussian cases,
  while B is the ratio between the effective bias, as defined in
  equation \ref{eq:beff}, in the Gaussian and non-Gaussian cases.
  Upper panels show the results for number counts and effective bias
  at two different redshifts, $z=0.5$ and $z=1$. Lower panels show
  results for the effective bias at $z=0.5$ for different wavenumbers,
  $k=0.1$, $0.01$ and $0.001$ $\MpcI$.}
\label{fig:ms}
\end{figure}

To better understand the reason for the weak dependence of the $\fnl$
constraints on the uncertain knowledge of the nuisance parameters, we
show in Figure \ref{fig:ms} by how much number counts and effective
bias change with respect to the value that they take in the Gaussian
case, as we vary the mass bias parameter $B_M$ (left panels) and the
intrinsic mass-scatter $\sigma_{\ln M}$ (right panels). As a reference
value for the non--Gaussianity, we take here $\fnl=10$, which is
comparable to the forecasted precision with which non--Gaussianity can
be constrained from our analysis. At $z=0.5$ the deviation of the
number counts from the non-Gaussianity (upper panels) varies only by
about one part over thousand when a generous range of variation is
allowed for both $B_M$ and $\sigma_{\ln M}$, with only a slighty
higher sensitivity to these parameters at $z=1$. In the bottom panels
of Figure \ref{fig:ms} we show the sensitivity of the effective bias on
nuisance parameters for different values of the wavenumber $k$. Results
are shown at $z=0.5$ which is close to the redshift where the effective
volume $V_{eff}$ reaches its maximum value (see left panel of
Figure \ref{fig:pesiz}). For the level of non--Gaussianity assumed here,
the deviation from the Gaussian effective bias is negligible at the
wavenumbers, $k\simeq 0.01\MpcI$, which are mostly weighted in the
computation of the Fisher Matrix (see right panel of
Figure \ref{fig:pesiz}). However, as expected, the effect of
non-Gaussianity on $b_{eff}$ shows up at very large scales, with a
deviation with respect to the Gaussian result by $\magcir 40$ per cent
for $k\simeq 10^{-3}\MpcI$. This highlights the importance for future
surveys to have a highly uniform calibration of the selection function
over large area of the sky, for them to be able to appreciate any
subtle scale dependence of the bias parameter. Also in this case, any
variation with the value of the nuisance parameters is far smaller
than the deviation from Gaussianity. This justifies the weak
dependence of the $\fnl$ constraints on the uncertain knowledge of the
cluster mass calibration.

\begin{figure}
\includegraphics[width=0.47\textwidth]{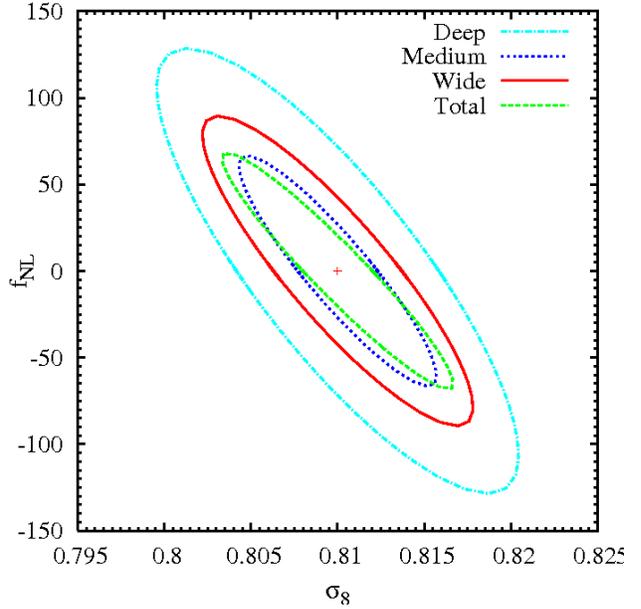}
\caption{Constraints at the 68 per cent confidence level on the
  $\fnl$--$\sigma_8$ plane for the {\em bright} subsamples of the
  three surveys, i.e. by including only clusters with fluxed 30 times
  larger than the detection flux. Dot-dashed
  cyan, dotted blue and solid red curves are for the Deep, Medium and Wide
  surveys, respectively, while the short-dashed green curve is for the
  combination of the three surveys. We assume here {\em strong prior} on the
  nuisance parameters and we include FM from Planck experiment.}
\label{fig:gold}
\end{figure}

As already mentioned in Sect. \ref{s:sur}, the {\em bright surveys}
contain only clusters for which a precise estimate of robust mass
proxies, such as $Y_X$ and $M_{gas}$, can be obtained. In this way,
the analysis of the {\em bright} subsamples offers us the possibility
of testing the trade-off between having smaller samples, for which
{\em strong prior} on the nuisance parameters can be assumed, and of
larger samples with a less certain knowledge of such parameters. In
Figure \ref{fig:gold} we show the constraints expected for the bright
subsamples of the three surveys. We note that this time the role of
the Medium and of the Wide surveys are reversed, with the former
providing the most stringent constraints. This is not surprising,
since the number of bright clusters in the Wide Survey falls below
that of the Medium Survey already at $z\simeq 0.1$ (see
Figure \ref{fig:nz}). Therefore, the richer statistics of distant
clusters in the Medium Survey more than compensates the more efficient
sampling of small-$k$ modes offered by the Wide survey. In terms of
figure-of-merit, we note that the resulting value from the combination
of the three bright surveys is $\fom \simeq 22.3$: this implies that
assuming strong priors for nuisance parameters in the bright surveys
has less constraining power than using {\em no priors} for the surveys
defined down to the detection flux limit.

\section{Discussion}

An interesting outcome of our analysis is the relative lack of
sensitivity of the forecasted $\fnl$ constraints on the nuisance
parameters: constraints on non--Gaussianity are tighter with larger
statistics, rather than with a smaller samples with better controlled
systematics in cluster mass estimates. While this is the case for the
purpose of constraining non--Gaussianity, it does not necessarily hold
for constraints on the Dark Energy equation of state. In order to
quantify the effect of an uncertain knowledge of the nuisance
parameters on Dark Energy constraints, we use as a reference the
redshift dependence of the equation of state, assumed e.g. in the Dark
Energy Task Force (DETF) report \citep[e.g.,][]{albrecht09},
$w(a)=w_0+(1-a) \, w_a$. Introducing the figure-of-merit
$\fomdetf=\left(\det\left[Cov(w_0,w_a)\right]\right)^{-1/2}$, we find
$\fomdetf\simeq 2066$, from the combination of mass function and
power spectrum, by adding the contribution of the three surveys and
assuming {\em strong prior} on the values of the nuisance
parameters. Such a high value drops to $\fomdetf\simeq 987$ under
the more realistic assumption of {\em weak priors} on the nuisance
parameters, while further reducing only by a small amount,
$\fomdetf\simeq 972$, if we assume no prior on such
parameters. Results on the constraining power of the high-sensitivity
X--ray surveys reported here on different Dark Energy models will be
presented in a future paper.

A first word of caution in the interpretation of such high values of
the DETF figure-of-merit lies in the assumption that scatter in the
mass bias is always assumed to be log-normal distributed. As discussed
by \cite{Shaw_etal09}, deviations from this assumption are in general
negligible as long as one relies on mass proxies which have a small
value of the scatter, while the uncertain knowledge of the
distribution of the scatter may significantly worsen constraints as the
scatter increases. In our samples of clusters identified down to the
detection limit, cluster selection is based on X--ray luminosity,
whose correlation with cluster mass can easily have a scatter as large
as 30 per cent. On the other hand, the effect should be much smaller
for the {\em bright} samples, that only contains clusters for which
measurements of robust low--scatter mass proxies are
available. Restricting our analysis to the {\em bright surveys}, we
find $\fomdetf=423$ and $176$ by assuming {\em strong priors} and
{\em no priors} on the nuisance parameters, respectively.  

Another word of caution lies in the assumption that redshifts for all
clusters in the surveys are perfectly known. An obvious objection when
discussing the cosmological exploitation of large surveys of clusters
concerns the possibility of measuring redshifts for all of them. While
discussing in detail the synergies between future X--ray and
optical/near--IR surveys is beyond the aim of this paper, it is worth
pointing out that future imaging and spectroscopic surveys, both
ground based (e.g., BigBOSS\footnote{http://bigboss.lbl.gov/},
PANSTARR\footnote{http://pan-starrs.ifa.hawaii.edu/},
LSST\footnote{http://www.lsst.org/}) and from space (e.g.,
EUCLID\footnote{http://arxiv.org/pdf/0912.0914},
JDEM\footnote{http://jdem.gsfc.nasa.gov/}) will provide either
photometric or spectroscopic redshifts for all clusters identified
within the WFXT surveys. Furthermore, all clusters in the {\em
  bright sample} will have a sufficiently large number of counts to
guarantee a rather precise X--ray spectroscopic redshift measurement
from the positions of metal lines \citep{giacconi09}. In general,
uncertainties in redshift measurements could be taken into account in
our analysis \citep[e.g.,][]{Cunha09}, at least as long as one has a
reliable estimation of redshift errors expected from different
observational techniques. While redshifts from optical
spectroscopy are generally so accurate that any uncertainty can be
neglected, the same is generally not true for photometric
redshifts. We also note that no attempt has been pursued so far to
assess in detail the reliability of measurements of clusters redshifts
from X--ray spectroscopy, as a function of signal-to-noise, cluster
temperature and redshift.

In our analysis we did not include the effect of redshift space
distortions in the distribution of galaxy clusters induced by peculiar
velocities \citep[e.g.,][]{KA87.1,White_etal09,Desjaques_Sheth10}. The
study of this effect to available galaxy redshift surveys
\citep[e.g.,][]{Guzzo_etal08} has indeed demonstrated that it provides
important constraints on the growth rate of density perturbations
\citep[see also][]{linder08}. Although we expect that its applications to
cluster surveys may be limited by the sparser sampling offered by the
cluster distribution, the effect of redshift-space distortions should
be in principle included when forecasting the cosmological
constraining power of future cluster surveys. We will present this
analysis in a forthcoming paper.

The analysis presented here demonstrates the potential that future
high--sensitivity X--ray cluster surveys could have in constraining
possible deviations from Gaussianity. However, it may be worth asking
what current data can tell us about such deviations. For instance, a
positive skewness has the effect of anticipating the first collapse of
massive DM halos. \citet{jee09} recently reported the discovery of an
unexpectedly massive galaxy cluster at $z\simeq 1.4$, XMMU-J2235.3,
identified as part of the initial 11 sq.deg. of the XMM
  Distant Cluster Project survey \citep{mullis05}, having a flux limit
  of $10\times 10^{-14}$ erg s$^{-1}$cm$^{-2}$. Based on weak lensing
\citep{jee09} and
  X-ray \citep{rosati09} analyses, a robust $1\sigma$ lower limit of
  $5\times 10^{14}$ is obtained for the cluster virial mass.  By
assuming a WMAP-5 cosmology, with $\sigma_8=0.81$ and $\Omega_m=0.28$,
and using the mass function by \citet{jenkins01}, \citet{jee09}
found that only $\simeq 5\times 10^{-3}$ of such massive clusters
should be expected within the survey area. Thus, they concluded that
XMMU-J2235.3 is a rather unlikely event in a standard cosmological
scenario. \citet{jimenez09} argued that, for a fixed value of
$\sigma_8$ ($=0.77$ in their analysis) the expected number of such
massive clusters can in fact be significantly enhanced in the case of
a positively skewed non-Gaussian distribution of primordial
perturbations.

\begin{figure}
\includegraphics[width=0.47\textwidth]{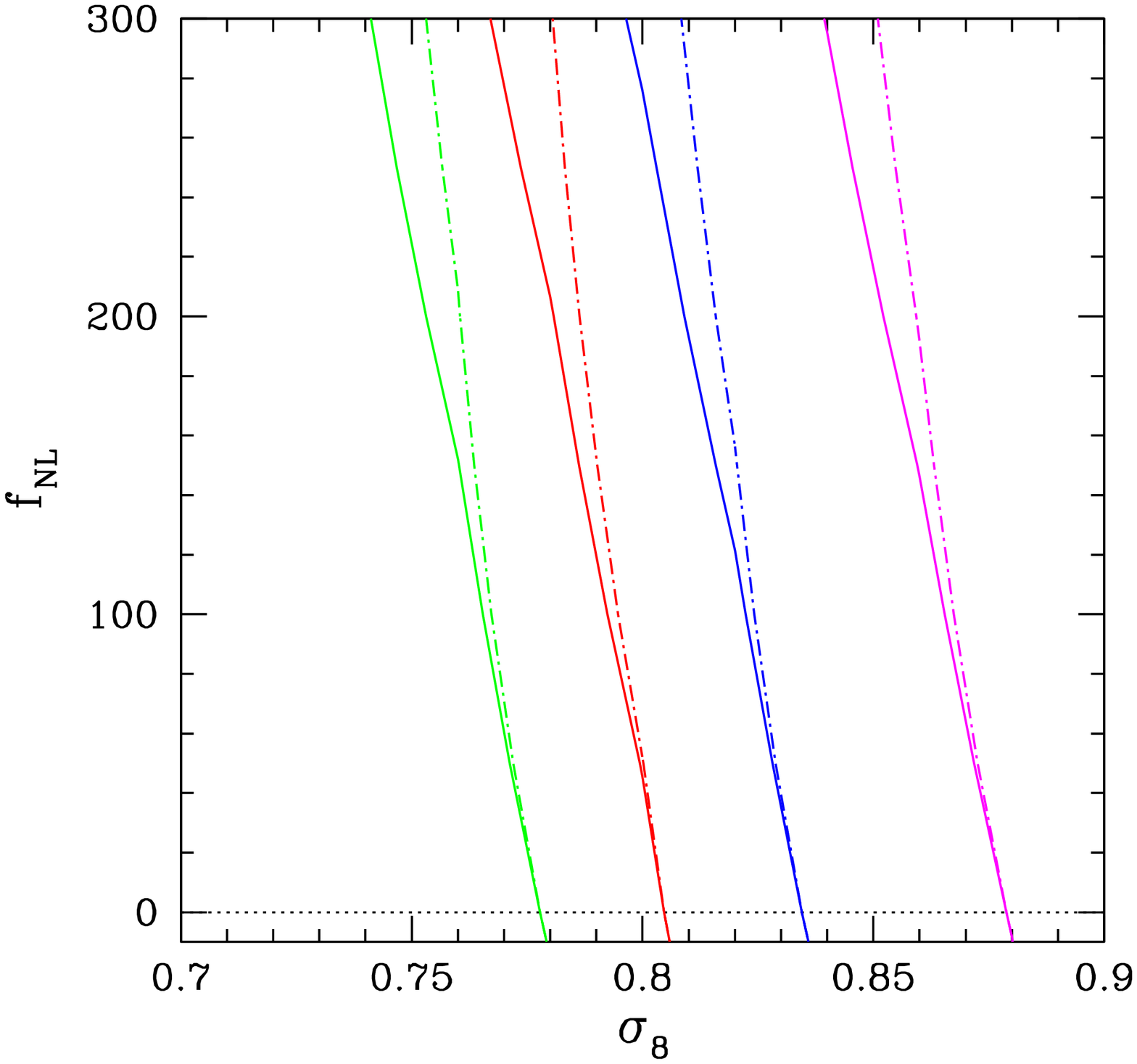}
\caption{The number of clusters with mass larger that $5\times
  10^{14}M_\odot$, found in the redshift range $1.4< z<2$ within the
  same survey area of 11 sq.deg. where the XMMU-J2235.3 cluster has
  been detected \protect\citep{jee09}. Non--Gaussian
mass function by \citet {loverde08} (dot--dashed curves) and by
\citet{matarrese00} (solid curves) are shown. From right to left,
  magenta, blue, red and green curves show the models on the
  $\sigma_8$--$\fnl$ plane predicting 0.05, 0.02, 0.01 and 0.005 clusters
  within the survey area, respectively. All other cosmological
  parameters have been kept fixed to our reference values (see
  text).}
\label{fig:xmm}
\end{figure}

In Figure \ref{fig:xmm} we show the curves in the $\sigma_8$--$\fnl$
plane corresponding to different numbers of clusters expected at
$z>1.4$ within 11 sq.deg. and having mass of at least $5\times
10^{14}M_\odot$. Results are given for the reference non--Gaussian
mass function from \cite{loverde08} (dot--dashed curves), that we used
for our forecasts, and for the mass function by \cite{matarrese00}
(solid curves). For both mass functions, we applied the correction to
$\Delta_c$ suggested by \cite{grossi09}. As previously discussed,
these two mass functions come from different approaches to approximate
the exact result for small values of $\fnl$. As expected, the
difference between the two mass functions becomes non negligible for
$\fnl >100$ for the rare event of such a massive cluster at $z\simeq
1.4$.  For each model, the four curves, from right to left, are for
0.05, 0.02, 0.01 and 0.005 such massive clusters found within the
survey area, respectively. For homogeneity with the analysis carried
out by \cite{jee09}, we used here the Gaussian mass function by
\cite{jenkins01}. While $\fnl$ and $\sigma_8$ are left free to vary,
all the other cosmological parameters are kept fixed at the fiducial
values adopted in our reference cosmological model (see
Sect. \ref{s:fm}). The results shown in this plot confirm that a
positive skewness helps increasing the expected number of
clusters. The effect of non-Gaussianity is strongly degenerate with
that of changing $\sigma_8$. For instance, increasing the expected
number of clusters by about a factor of ten for $\sigma_8=0.8$
requires $\fnl$ values in excess
of the range allowed already at present by CMB \citep[e.g.,][ and
references therein]{komatsu10} and Large Scale Structure (LSS) \citep{SL08.1} data. On the
other hand, the same boost in the cluster number can be achieved by
requiring $\fnl\simeq 100$ and increasing $\sigma_8$ to $\simeq 0.87$,
again in tension with current CMB and LSS constraints. The
conclusion of this analysis is that for XMMU-J2235.3 not to be a very
unlikely event, a degree of non--Gaussianity in excess of the
currently allowed CMB bounds is required, unless one wants to violate
current constraints on $\sigma_8$. Clearly, more than a single
detection of such massive distant clusters are needed to draw firm
conclusions. However, this example further confirms the strong
constraining power of even few massive clusters at $z>1$. In addition,
since galaxy clusters probe much smaller scales than the CMB, they
offer a complementary approach to test a possible scale--dependence of
non--Gaussianity.

\section{Conclusions}
We presented forecasts on the capability of future high-sensitivity
X--ray surveys of galaxy clusters to provide constraints on deviations
from Gaussian primordial perturbations. Our analysis is based on
computing the Fisher Matrix (FM) for the information given by the
evolution of mass function and power spectrum of galaxy
clusters. Following the approach by \citet{tegmark97} to compute the
power-spectrum FM \citep[see also][]{majumdar04,rassat08,stril09}, we
include in the analysis the information related to the possible
scale--dependence of the linear bias, which represents a unique
fingerprint of non-Gaussianity \citep[e.g.,][and references
therein]{verde10}. According to the self--calibration
approach, the model parameters entering in the FM estimate are nine
cosmological parameters and 4 nuisance parameters, the latter defining
the relation between cluster mass and observable upon which cluster
selection is based. Our analysis is based on assuming an observational
strategy designed for the Wide Field X--ray Telescope (WFXT,
e.g. \citealt{giacconi09}), in which a {\em Wide} Survey covering most
part of the extragalactic sky is complemented by a {\em Medium} and
by a {\em Deep} Survey (see Table \ref{t:sur}). The latter provides
mass proxies down to the flux limit for cluster identification in the
Wide Survey (see Table 1). We showed forecasts for the two parameters
that, for a fixed expansion history, define the timing of cosmic
structure formation, namely $\sigma_8$ and $\fnl$, while marginalizing
over all the remaining parameters. Informations on such constraints are
quantified by introducing the figure-of-merit for structure formation
timing of Eq.(\ref{eq:fom}).

The main results obtained from our analysis can be summarized as follows.
\begin{description}
\item[(a)] Power spectrum and number counts of galaxy clusters are
  highly complementary in providing constraints: while the former is
  sensitive to deviations from Gaussianity, through the scale
  dependence of the bias, the latter is mostly sensitive to
  $\sigma_8$.
\item[(b)] Most of the constraining power for these two parameters
  lies in the {\em Wide} Survey, while the {\em Medium} and the {\em
    Deep} Surveys play an important role for the estimate of X--ray
  mass proxies for $\simeq 2\times 10^4$ clusters out to $z\sim 1.5$.
\item[(c)] Combining number counts and power spectrum information for
  the three surveys turns into $\Delta \fnl\simeq 10$ for the
  $1\sigma$ uncertainty with which a deviation from Gaussianity
  associated to a ``local shape'' model can be
  constrained. Correspondingly, we find $\fom
  \simeq 39$ for the figure-of-merit of structure formation timing.
\item[(d)] Quite interestingly, while the value of $\fom$
  significantly worsens when assuming more
  conservative priors on the nuisance parameters, the above constraint
  on $\fnl$ is weakly sensitive on such priors.
\item[(e)] The presence of a cluster as massive as XMMU-J2235.3 at
  $z\simeq 1.4$ \citep{jee09} turns out to be a rather unlikely event,
  even allowing for an amount of non-Gaussianity consistent with
  current CMB \citep[e.g.,][]{komatsu10} and LSS \citep{SL08.1}
  constraints. This further demonstrates the strong constraining power
  of detecting an even small number of massive high-$z$ clusters.
\end{description}

Our analysis lends support to the important role that future cluster
surveys will play in constraining deviations from the Gaussian
paradigm, with far reaching implications on the primordial mechanisms
which seeded density inhomogeneities. The reliability of our
forecasts relies on the possibility of calibrating to high precision a
universal expression for mass function and large--scale bias. A number of independent groups
\citep{grossi07,DA08.1,DE08.1,grossi09,Giannatonio_Porciani09,pillepich10}
carried out large N--body simulations with non--Gaussian initial
conditions, finding in general a quite good agreement for the
calibration of both mass function and bias. However, the precision
required for the calibration of such quantities, for them not to spoil
the constraining power of large surveys, is probably higher than what
reached at present. First assessments of the impact of uncertainties
in the mass function calibration on DE constraints have been already
presented \citep[e.g.][]{wu09} and indicate that such
uncertainties may not be negligible. There is no doubt that larger
suites of non--Gaussian simulations are required to calibrate mass
function and large-scale bias also for a range of models beyond the
local non--Gaussian models that we considered in the analysis
presented here.

A few days after our paper, \citet{cunha10}
  also subbitted a paper regarding the study of constraints on
  non-Gaussian parameter $f_{NL}$ from cluster surveys. They use the
  Fisher matrix approach applied to the count-in-cell method to
  extract information on evolution of cluster number density and
  clustering. Differently from the previous paper by \cite{oguri09},
  \citet{cunha10} also included the contribution from covariance
  between counts in different cells. This allowed them to sample the
  power spectrum over a large scale range, thus obtaining tighter
  constraints on $f_{NL}$ than \cite{oguri09}. By specialising their
  analysis for the mass selection and sky coverage expected for the
  DES\footnote{https://www.darkenergysurvey.org/} optical survey, they forecast a precision of
  $\sigma_{f_{NL}}\simeq 1$--5. A detailed comparison between our and
  their analysis is not straightforward.  Besides using different
  survey specifications, our and their analyses also uses different
  prescriptions for the mass function and the bias. Just as an example
  of how sensitive the choice of the bias model is, we verified that
  excluding the bias correction suggested by \citet{grossi09} (the
  factor $q=0.75$ that we introduce after Eq.\ref{eq:bias}), the expected errors in $f_{NL}$ would decrease by approximately a factor of 2. This further emphasizes
  the need for a precise calibration of the model mass function and
  bias through extended sets of non-Gaussian N--body simulations.

\section*{Acknowledgments.} 
We acknowledge useful discussions with Cristiano Porciani, Anais
Rassat and Paolo Tozzi. We thank Elena Pierpaoli, Ravi Sheth and Licia
Verde for useful comments on the first version of this paper. This
work has been partially supported by the PRIN-MIUR Grant ``The Cosmic
Cycle of Baryons'', by a ASI-AAE Theory Grant, by the ASI Contract
No. I/016/07/0 COFIS, the ASI/INAF Agreement I/072/09/0 for the Planck
LFI Activity of Phase E2, the ASI/I/023/05/0 and the ASI/I/088/06/0

\bibliographystyle{mn2e}
\bibliography{sartoris}

\end{document}